%% file: main.tex
\documentclass[superscriptaddress,longbibliography,nofootinbib,twocolumn,notitlepage,10pt]{revtex4-1}
\usepackage{amsthm}
\usepackage{amsmath}
\usepackage{bbold}
\usepackage{bbm}
\usepackage{graphicx}
\usepackage{hyperref}
\usepackage{listings}
\usepackage{mathtools}
\usepackage{float}
\usepackage{stmaryrd}
\usepackage{algorithm}
\usepackage{algpseudocode}
\usepackage{makecell}
\usepackage{multirow}
\usepackage{physics}
\usepackage{wasysym}
\usepackage{bm}
\usepackage[dvipsnames]{xcolor}

\hypersetup{colorlinks,citecolor=blue,linkcolor=blue,urlcolor=blue}

\newcommand{\Google}{\affiliation{
Google Quantum AI, Venice, CA 90291, USA}}

\newcommand{\GoogleSEA}{\affiliation{
Google, 601 N 34th St, Seattle, WA 98103, USA}}

\begin{document}

\title{Efficient approximation of experimental Gaussian boson sampling}

\date{\today}

\author{Benjamin Villalonga}
\Google
\author{Murphy Yuezhen Niu}
\Google
\author{Li Li}
\affiliation{Google Research, 1600 Amphitheatre Parkway, Mountain View, California 94043, USA}
\author{Hartmut Neven}
\Google
\author{John C. Platt}
\GoogleSEA
\author{Vadim N. Smelyanskiy}
\Google
\author{Sergio Boixo}
\email[Corresponding author: ]{boixo@google.com}
\Google

\begin{abstract}
  Two recent landmark experiments have performed Gaussian boson sampling (GBS) with a non-programmable linear interferometer and threshold detectors on up to 144 output modes (see Refs.~\onlinecite{zhong_quantum_2020,zhong2021phase}).
  Here we give classical sampling algorithms with better total variation distance and Kullback-Leibler divergence than these experiments and a computational cost quadratic in the number of modes.
  Our method samples from a distribution that approximates the single-mode and two-mode ideal marginals of the given Gaussian boson sampler, which are calculated efficiently.
  One implementation sets the parameters of a Boltzmann machine from the calculated marginals using a mean field solution.
  This is a 2nd order approximation, with the uniform and thermal approximations corresponding to the 0th and 1st order, respectively.
  The $k$th order approximation reproduces Ursell functions (also known as connected correlations) up to order $k$ with a cost exponential in $k$ and high precision, while the experiment exhibits higher order Ursell functions with lower precision.
  This methodology, like other polynomial approximations introduced previously, does not apply to random circuit sampling because the $k$th order approximation would simply result in the uniform distribution, in contrast to GBS.
\end{abstract}
\maketitle

\section{Introduction}
\label{sec:intro}
\input{intro}

\section{Classical mockup sampling methods}
\label{sec:classical}
\input{classical}

\section{Numerical results}
\label{sec:results}
\input{results}

\section{Discussion}
\label{sec:discussion}
\input{discussion}

\begin{acknowledgements}
We thank  Daniel Eppens, Sergei Isakov, and Wojtek Mruczkiewicz for their help with running verification jobs on Google servers.
We are also thankful to Scott Aaronson, Ish Dhand, Joshua V. Dillon, Gil Kalai, Seth Lloyd, Chao-Yang Lu, John Martinis, and Jelmer Renema for interesting and helpful discussions.
\end{acknowledgements}

\appendix
\input{appendix}

\newpage
\bibliography{BosonSampling}

\end{document}

%% file: intro.tex
Quantum computers hold the promise of efficiently solving certain computational tasks that are beyond the capabilities of classical computers.
There is still a long path ahead towards the realization of a large-scale, error-corrected, programmable quantum computer.
Nevertheless, in 2019 Ref.~\onlinecite{arute_quantum_2019} reported a beyond-classical computation through the task of random circuit sampling (RCS)~\cite{boixo_characterizing_2018} using a fully programmable quantum processor; this announcement has been followed by similar experiments~\cite{wu2021strong,zhu2021quantum}. 
RCS uses standard quantum circuits and there is a substantial body of literature studying RCS in complexity theory~\cite{boixo_characterizing_2018,aaronson2017complexity,bouland2019complexity,movassagh2019quantum,aaronson2019classical,bouland2021noise,kondo2021fine} and computational methods~\cite{haner_0.5_2017,boixo2017simulation,villalonga2020establishing,gray_hyper-optimized_2020,zhang_alibaba_2019,huang2020classical,de_raedt_massively_2018,li_quantum_2018,chen_64-qubit_2018,chen_classical_2018,chen_quantum_2019}, including approximations~\cite{boixo_fourier_2017,markov2018quantum,villalonga_flexible_2019,napp_efficient_2019,noh_efficient_2020,barak_spoofing_2020,zhou2020limits,bravyi2021classical,pan_simulating_2021}.
As of today, despite substantial improvements in classical algorithms and implementations~\cite{boixo2017simulation,villalonga2020establishing,gray_hyper-optimized_2020,zhang_alibaba_2019,huang2020classical}, the sampling tasks reported in Refs.~\cite{arute_quantum_2019,wu2021strong,zhu2021quantum} have not been reproduced with classical supercomputers. 

\begin{center}
\begin{table}
\begin{tabular}{ |c||c|c|c|c|c| } 
\hline
\textbf{Dataset} & experiment & $N$ & waist ($\mu$m) & $P (W)$ & \thead{theoretical \\ mean \\ click num.} \\
\hline
\hline
\textbf{1} & 1 (Ref.~\onlinecite{zhong_quantum_2020}) & 100 & - & - & 41.04 \\ 
\hline
\textbf{2.a.1} & \multirow{7}{*}{2 (Ref.~\onlinecite{zhong2021phase})} & \multirow{7}{*}{144} & \multirow{2}{*}{125} & 0.5 & 7.27 \\ 
\textbf{2.a.2} & & & & 1.412 & 19.26 \\ 
\cline{1-1} \cline{4-6}
\textbf{2.b.1} & & & \multirow{5}{*}{65} & 0.15 & 5.98 \\ 
\textbf{2.b.2} & & & & 0.3 & 11.94 \\ 
\textbf{2.b.3} & & & & 0.6 & 24.66 \\ 
\textbf{2.b.4} & & & & 1.0 & 41.79 \\ 
\textbf{2.b.5} & & & & 1.65 & 66.87 \\ 
\hline
\end{tabular}
\caption{
\label{tab:datasets}
Experimental datasets of Refs.~\onlinecite{zhong_quantum_2020,zhong2021phase}.
As explained by the authors, reducing the focus waist or increasing the power $P(W)$ of the  pump results in an increased mean click number in the output.
See Figs.~\ref{fig:hist_click_number}, \ref{fig:click_number_dist} and~\ref{fig:moment_difference} and App.~\ref{app:moments} for details on the distributions of click number and their moments.
}
\end{table}
\end{center}

In 2020, Ref.~\onlinecite{zhong_quantum_2020} reported a landmark Gaussian boson sampling experiment (GBS)~\cite{hamilton_gaussian_2017,quesada_gaussian_2018,gupt_classical_2018,bjorklund_faster_2019,kruse_detailed_2019,zhong2019experimental,wu_speedup_2019,quesada_quadratic_2020,quesada_exact_2020,drummond2021simulating,li_benchmarking_2020} in a photonic interferometer with 50 input single-mode squeezed states and threshold detectors on 100 output modes, followed by Ref.~\onlinecite{zhong2021phase} with 144 output modes and improved calibration.
The linear interferometers used in these experiments are not programmable.
The cost of calculating the ideal output probability of a given bit string is exponential in the number of 1s or detector {\it clicks}. The mean number of clicks is as high as 66.87 in dataset {\bf 2.b.5}, see Table~\ref{tab:datasets}.
Refs.~\onlinecite{zhong_quantum_2020,zhong2021phase} show that some known mockup distributions are further from the ground truth or ideal distribution than the experiment.
The mockup distributions considered are: uniform samples, distinguishable bosons, and a thermal approximation. Ref.~\onlinecite{zhong2021phase} also shows that high order Ursell functions can be detected in the experiment. 

The only known efficient general approximation of the RCS output is the uniform distribution over bit strings~\cite{boixo_characterizing_2018,boixo_fourier_2017}.\footnote{Ref.~\onlinecite{zhou2020limits} gives an efficient approximation better than uniform for one dimensional random circuits with gate fidelity below some threshold.} Indeed, the marginal probabilities for any subset of qubits are also exponentially close to uniform due to the highly entangled nature of the RCS output. The difficulty of approximating the RCS output distribution can be appreciated with the observation that even a single discrete error on a random quantum circuit will result  in an output distribution uncorrelated with the ground truth~\cite{boixo_characterizing_2018}. Furthermore, such high error sensitivity of RCS allows us to use it as an estimator of system fidelity~\cite{boixo_characterizing_2018,neill2018blueprint,arute_quantum_2019,liu2021benchmarking}. 

The situation is very different for boson sampling: marginals of the output distribution are far from uniform~\cite{aaronson2011computational,aaronson2013bosonsampling,kalai_gaussian_2014,clifford2018classical,ivanov2019complexity,renema_marginal_2020}. This makes it harder than in the RCS case to reach a conclusion about the computational hardness of a given experiment as a sampling problem, and GBS does not result in an estimate of fidelity. Note that GBS is based on the continuous variable formulation of bosonic interference~\cite{weedbrook2012gaussian,serafini2017quantum}, which is by design not as sensitive to discrete errors such as losing photons in a specific mode. Furthermore, in GBS photon loss is incorporated in the ground truth, as it is included in the quantum continuous variable description of the experiment. 

Consequently, there exist polynomial approximations of boson sampling~\cite{kalai_gaussian_2014,rahimi2016sufficient,qi_regimes_2020,renema2017efficient,renema2018classical,renema_simulability_2020,renema_sample-efficient_2020,renema_marginal_2020}. One type of approximation aims at finding a positive quasi-probability description of the experiment from first principles, but it does not apply to the parameter regime~\cite{rahimi2016sufficient,qi_regimes_2020} of the experiments of Refs.~\onlinecite{zhong_quantum_2020,zhong2021phase}.\footnote{The dark count rate of the superconducting nano-wire single-photon detector used is very low, $p_D \sim 10^{-4}$~\cite{qi_regimes_2020}.} A different kind of approximation was first developed as a polynomial approximation of noisy permanents in the context of boson sampling with a well defined number of photons~\cite{kalai_gaussian_2014,renema2017efficient,renema2018classical}. Although it can be extended to GBS, which uses quantum continuous variables~\cite{renema_simulability_2020,renema_marginal_2020}, it might require relatively expensive high order polynomial calculations~\cite{zhong_quantum_2020,renema_sample-efficient_2020}.

In this paper we give a different approximation to boson sampling and we show that the GBS task of Refs.~\onlinecite{zhong_quantum_2020,zhong2021phase} can be approximated with better statistical distance than the experiment efficiently, with a cost quadratic in the number of modes.
We proposed the basic idea, related to previous  polynomial approximations~\cite{kalai_gaussian_2014,renema_marginal_2020} but more directly applicable to GBS with continuous variables, the same day that Ref.~\onlinecite{zhong_quantum_2020} appeared~\cite{aaronson_blog_2020}.
The starting point is that the ideal two-mode marginals, or two-mode correlations, are easy to compute both for standard boson sampling~\cite{ivanov2019complexity,clifford_classical_2017,clifford2018classical,aaronson2013bosonsampling,renema_marginal_2020} and GBS~\cite{zhong_quantum_2020,serafini2017quantum}. 
Our method samples from a distribution that approximates all the single-mode and two-mode marginals.
We implement two heuristic algorithms that achieve that.
The first one sets the parameters of a Boltzmann machine using the calculated correlations to compute the effective, mean field coupling constants of a fully-connected Ising model.
The main difficulty is to compare with the ground truth, given that calculating the ideal probabilities corresponding to the experiment is expensive (see App.~\ref{app:probabilities}).
Note that the uniform distribution is a 0th order approximation in this method, and a thermal approximation is the 1st order approximation.
This method is capable of generating millions of bit strings per minute on a single workstation. 
Our methodology can be extended to order $k$ approximations with a cost exponential in $k$, and it does not capture Ursell functions (also known as connected correlations) beyond $k$th order.
Note that Ref.~\onlinecite{bulmer2021boundary} has recently introduced substantial improvements for the exact calculation of the ground truth probabilities, although the cost is still exponential and impractical for the exact simulation of the GBS experiments of Refs.~\onlinecite{zhong_quantum_2020,zhong2021phase}.
The same reference also introduced an approximate and efficient sampling algorithm that performs better than a thermal sampler.

\begin{figure}[t]
\centering
\includegraphics[width=1.00\columnwidth]{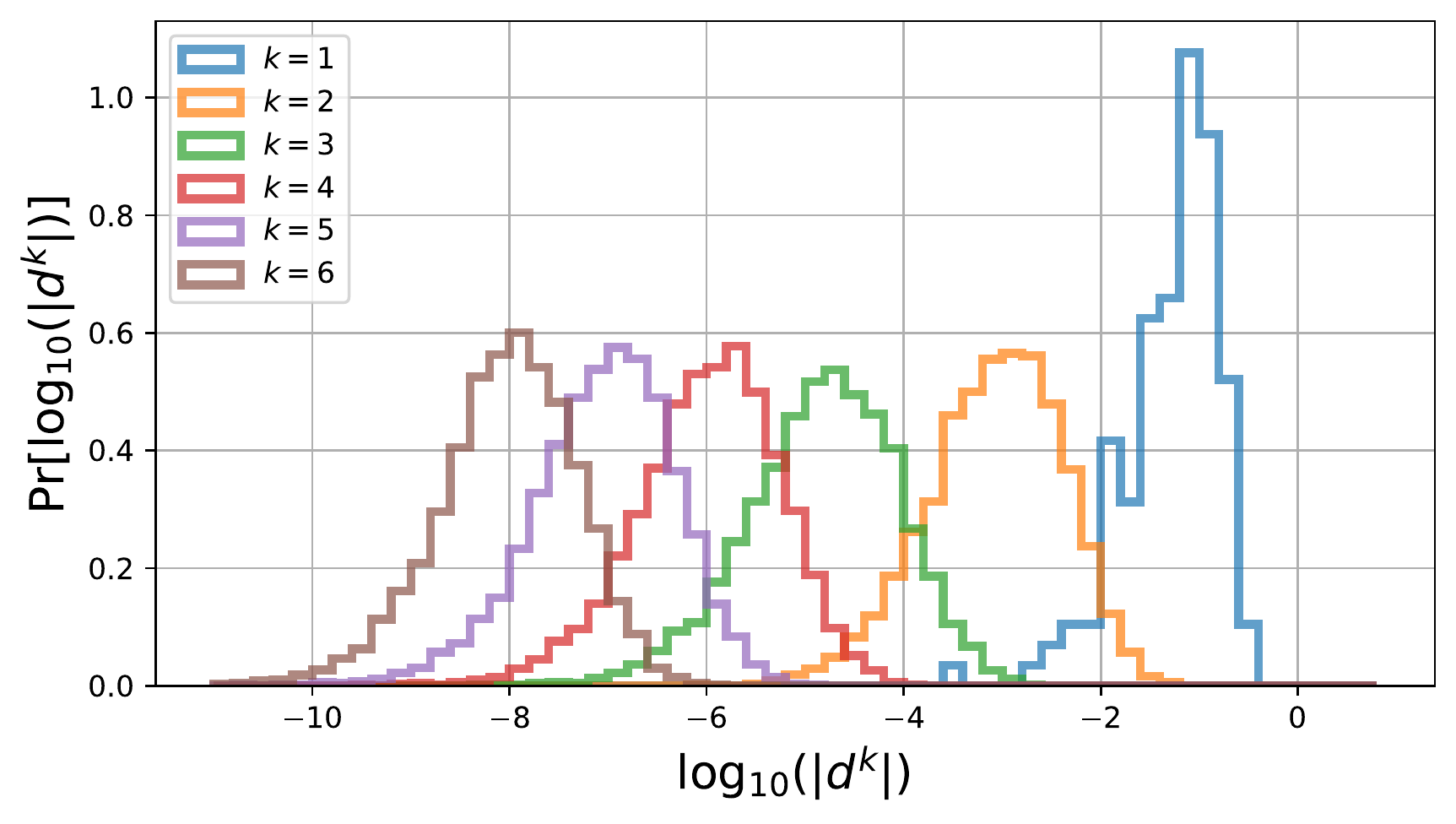}
\caption{\label{fig:correlations} 
Histogram of the logarithm of the absolute value of the $k$th order Ursell functions $|d^k|$ (see App.~\ref{app:connected_correlations}) of the ground truth (ideal distribution) for the GBS experiment in Ref.~\onlinecite{zhong2021phase} for orders $k = 1, \ldots, 6$ (dataset \textbf{2.b.5}, see Table~\ref{tab:datasets}).
Order 1 refers to the difference between one-mode marginal probabilities and $1 \over 2$; order 2 refers to the Ursell functions $d^2 = \left\langle z_i z_j \right\rangle - \left\langle z_i \right\rangle \left\langle z_j \right\rangle$; order $k$ Ursell functions $d^k$ are generalized as in App.~\ref{app:connected_correlations}.
We find numerically that typical Ursell functions decay exponentially with order $k$.
For each order $k$, we include data from up to 10000 randomly selected subsets of $k$ modes (for $k = 2$ we use all ${144 \choose 2} = 10296$ pairs of modes).
} 
\end{figure}

Recently, Ref.~\onlinecite{popova_cracking_2021} proposed an alternative heuristic to estimate a given output probability with a $k$th order polynomial and multiplicative error. The required order of their approximation will increase with the number of clicks and the calculation needs to be performed repeatedly for each output bit string. This is in contrast with our method, as we do not estimate any global probabilities. 
While the approximation of Ref.~\onlinecite{popova_cracking_2021} is in some sense exponential in the number of clicks, the authors argue that a fourth order approximation would suffice to reproduce the 100 mode experiment in Ref.~\onlinecite{zhong_quantum_2020} based on numerical studies with 30 modes.

%% file: classical.tex
The output of a GBS experiment with threshold detectors over $N$ modes is a bit string $\mathbf{z}$ of length $N$.
Computing a bit string probability exactly is exponentially expensive ---\#P-hard, indeed~\cite{hamilton_gaussian_2017,quesada_gaussian_2018}--- in the number of clicks (number of ones or photons detected by the threshold detectors).
Nevertheless, computing probabilities marginalized over subsets with a few modes is efficient (see App.~\ref{app:probabilities}).
More explicitly, the cost is exponential in the number of clicks, which is now upper bounded by the size of the subset of modes, and hence small.

In this section we describe a family of mockup samplers which avoid computing global probabilities of bit strings.
These samplers aim at sampling from a distribution with correct marginals up to $k$th order.
In particular, we say that a sampler is of order $k$ if it approximates the marginal probability distributions of subsets of at most $k$ modes of the ground truth (ideal GBS).
In Sections~\ref{sec:boltzmann} and~\ref{sec:greedy} we present two methods to instantiate $k$th order samplers.
In Section~\ref{sec:results} we compare the performance of these samplers against the experiments of Ref.~\onlinecite{zhong_quantum_2020,zhong2021phase}.

\begin{figure}[t]
\centering
\includegraphics[width=1.00\columnwidth]{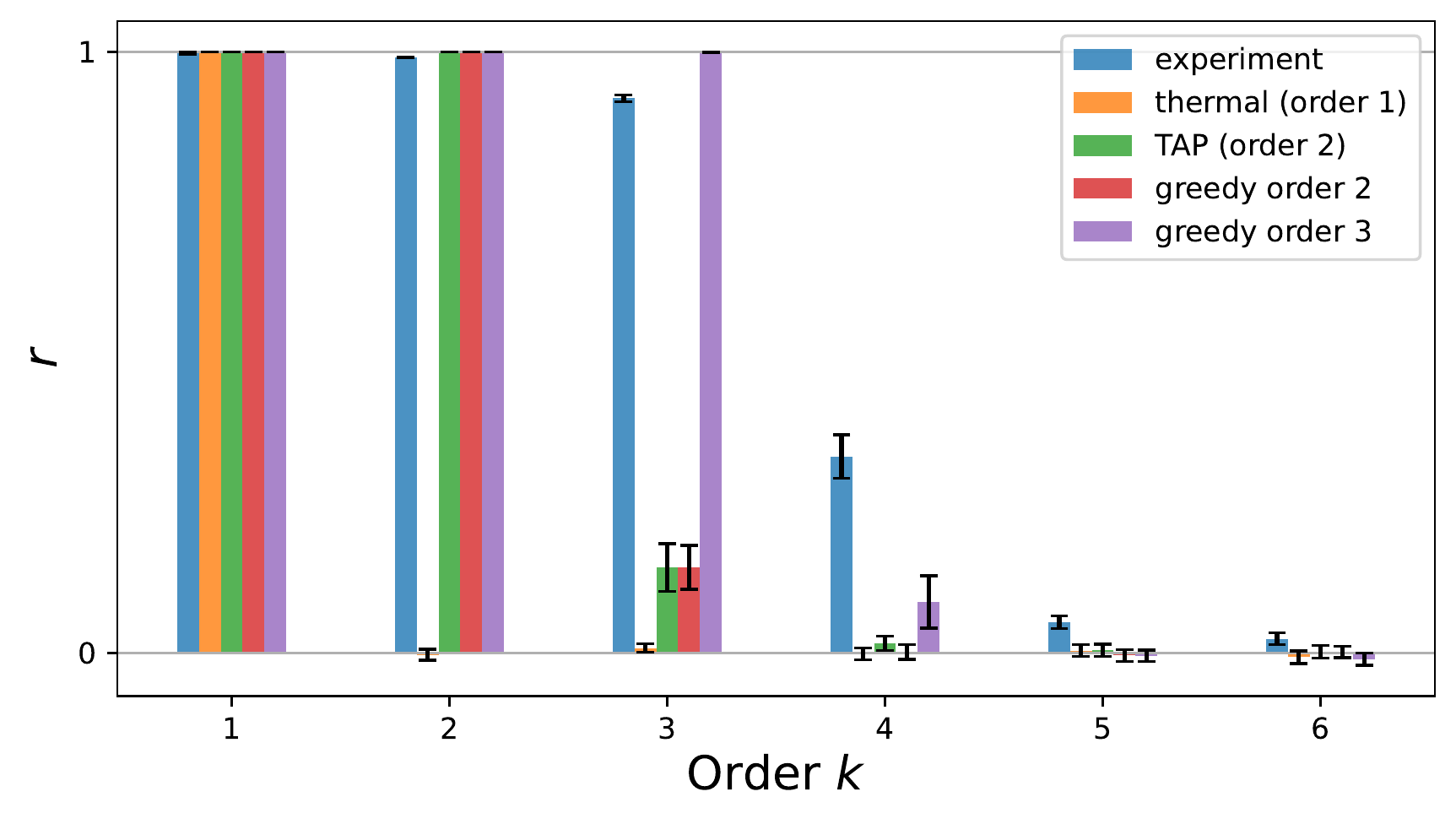}
\caption{\label{fig:pearson} 
Pearson correlation coefficient $r$ between the ideal values of the Ursell functions and their empirical counterparts for the different samplers, for orders $k = 1, \ldots, 6$ and for dataset \textbf{2.b.5}.
For each order $k$, up to 10000 randomly chosen subsets of modes were considered; in the case of $k = 1$ and 2 we used all $N = 144$ and ${N \choose 2} = 10296$ subsets, respectively.
Error bars correspond to the standard deviation over 500 bootstrapping resamples.
The empirical values are computed over a set of 10 million bit strings.
}
\end{figure}

\subsection{Boltzmann machines}
\label{sec:boltzmann}
The expected best $k$th order approximation corresponds, from the maximum entropy principle, to a Boltzmann machine (BM) (see App.~\ref{app:max_entropy}): 
\begin{align}
\label{eq:bm}
    p(\mathbf{z}) = \frac{1}{Z} \exp \left(\sum_{a} \lambda_a z_a + \sum_{a < b} \lambda_{a, b} z_a z_b \right. \nonumber \\
    \left.+ \sum_{a < b < c} \lambda_{a, b, c} z_a z_b z_c + \ldots \right) \text{,}
\end{align}
where $Z$ represents the partition function, which normalizes the probability distribution.

Trivially we consider the uniform distribution over bit strings as the 0th order approximation.
The 1st order approximation samples each bit independently with its correct average, and therefore uses only one-mode marginals; this corresponds to a BM with only the first summand in the exponent of Eq.~\eqref{eq:bm}.
We call this the thermal approximation.
Consistent with the results reported in Ref.~\onlinecite{zhong_quantum_2020}, this approximation performs worse than the experiment (see Section~\ref{sec:results}).\footnote{Ref.~\onlinecite{zhong_quantum_2020} uses a different thermal approximation. They use thermal states as input to the linear interferometer, while we approximate the output of the linear interferometer as a thermal state. The results are nevertheless similar.}

At order $k\geq2$ we can train such a BM, which includes the first $k$ summands in the exponent of Eq.~\eqref{eq:bm}, through gradient descent of the log-likelihood~\cite{nguyen_inverse_2017}.
Estimating the gradient of the log-likelihood requires estimating the correlations of the ground truth (ideal Gaussian boson sampling), and of the BM.
While the ground truth correlations can be calculated directly with machine precision (see App.~\ref{app:probabilities}), the correlations of a large fully connected BM can only be estimated through sampling.~\footnote{Note that this is not just a problem with estimating the partition function, which might be addressed with the pseudo-likelihood method.
The problem is that we do not have samples from the ground truth, but only the marginal probabilities.} 
Therefore, training a BM requires order $\frac{1}{g^2}$ samples of the intermediate BM per training step, where $g$ is the required precision in the marginal probabilities.
Given the exponentially decreasing values of the correlations with $k$ (see Fig.~\ref{fig:correlations}), this training requires exponentially increasing number of samples in $k$ to achieve a fixed relative error in the order of the correlations considered. 

Given the poor scaling of the gradient descent method to train the BM with the desired precision, we choose to find the parameters for the 2nd order BM through a mean field approximation, which avoids sampling the BM during the training.
In order to do so, it is more natural to rewrite the BM in terms of spin variables, $\mathbf{s} = \{s_a\}_{a = 1}^N$, as opposed to the Boolean variables of Eq.~\eqref{eq:bm}.
This is achieved by the change of variables $s_a = 2z_a - 1$.
In this language, the probability of a spin string $\mathbf{s}$ is:
\begin{align}
\label{eq:p_spin}
    p(\mathbf{s}) = \frac{1}{Z} \exp \left[ -H(\mathbf{s}) \right] \text{,}
\end{align}
where $H(\mathbf{s})$ is the fully-connected Ising Hamiltonian:
\begin{align}
\label{eq:ising}
    H(\mathbf{s}) = -\sum_a h_a s_a - \sum_{a < b} J_{a, b} s_a s_b \text{.}
\end{align}

In order to find the coupling constants of Eqs.~\eqref{eq:p_spin} and~\eqref{eq:ising} we now use the Thouless, Anderson, Palmer (TAP) mean field approximation, which yields~\cite{thouless1977solution,nguyen_inverse_2017}:
\begin{align}
    \label{eq:tap_constants_a}
    J^{\rm TAP}_{a, b} = &\frac{-2 (C^{-1})_{a, b}}{1 + \sqrt{1 - 8 (C^{-1})_{a, b} \left\langle s_a \right\rangle \left\langle s_b \right\rangle}} \\
    \label{eq:tap_constants_b}
    h^{\rm TAP}_a = &- \sum_{a \neq b} (J^{\rm TAP}_{a, b})^2 (1 - \left\langle s_b \right\rangle^2) \nonumber \\
    &- \sum_{a \neq b} J^{\rm TAP}_{a, b} \left\langle s_b \right\rangle + {\rm arctanh} (\left\langle s_a \right\rangle) \text{.}
\end{align}
Note that the expressions in Eqs.~\eqref{eq:tap_constants_a} and \eqref{eq:tap_constants_b} are a function of the one-spin magnetizations $\left\langle s_a \right\rangle$ (one-mode marginals) and the covariance matrix of the spins $C$ (two-mode marginals). Other mean field solutions seem to give similar results.

The next step is using this BM to produce mockup samples.
We do this using standard Gibbs sampling~\cite{gelfand}, as the probability of one bit (spin) conditional on all the others is easy to calculate (see App.~\ref{app:gibbs}). 
Note that this algorithm runs in $\mathcal{O}(N^2 L)$ time, where $N$ is the number of modes and $L$ the number of samples.
The numerical results of this method are discussed in Section~\ref{sec:results}.

\subsection{Greedy heuristic for generating bit strings with desired $k$th order marginals}
\label{sec:greedy}

In this section we describe an alternative greedy heuristic to generate a set of $L$ $N$-bit strings with approximately correct marginal probabilities. We encode the set of bit strings in a matrix $S$ of size $L \times N$ with entries either 0 or 1.
Each row corresponds to a bit string and each column to one of the $N$ modes.
Our goal is to choose each entry of the matrix to be either a 0 or a 1 in a way such that the empirical marginals up to order $k$ of the set of $L$ bit strings are as close as possible to the theoretical marginals.

We initially work with the first $k$ columns of matrix $S$ and iterate over all $L$ rows, where $k$ is the order of the approximation.
On iteration $i$ we choose the bit string of bits $S_{i,1}$ through $S_{i, k}$ to complete the sub-matrix of $S$ with rows from 1 through $i$ and columns 1 through $k$, which we denote by $S_{1:i, 1:k}$.
We choose the $k$-bit string $S_{i, 1:k}$ which minimizes the $\ell_1$ distance between the vectors of empirical and theoretical (ideal) marginal probabilities for the first $k$ modes.
After $L$ iterations we have placed all matrix elements in sub-matrix $S_{1:L, 1:k}$.
Finally we shuffle all rows before proceeding to the next column.

We now place bits on the ($k+1$)th column through the $N$th column of $S$.
For each column $j$, with $k + 1 \leq j \leq N$, we iterate over rows.
On iteration $i$ we choose the bit $S_{i, j}$ such that it minimizes the $\ell_1$ distance between the vectors of empirical and the theoretical marginal probability distributions of order $k$ that involve mode $j$ and $k - 1$ modes from the set $\{ l\}_{l = 1}^{j}$.
After $L$ iterations we have placed all matrix elements in column $j$ of $S$.
We now shuffle all rows of sub-matrix $S_{1:j, 1:L}$.
We repeat this procedure sequentially over all columns.

We see numerically that the $L$ bit strings thus generated produce a vector of empirical probability distributions with $\ell_1$ distance of order $\mathcal{O}(1/L)$ to the theoretical marginals, i.e., of the order of the rounding error.
This algorithm runs in $\mathcal{O}(N^k 2^k L)$ time.

Note that the samples generated this way are not i.i.d.
We can reduce the correlations between them by randomly selecting a subset of them for the output.
We can also generate i.i.d. samples by repeating this method many times and randomly selecting a single bit string per run.
In addition, one might try variations of this algorithm, e.g. iterating over rows and columns in different orders. 

\begin{figure}[t]
\centering
\includegraphics[width=1.00\columnwidth]{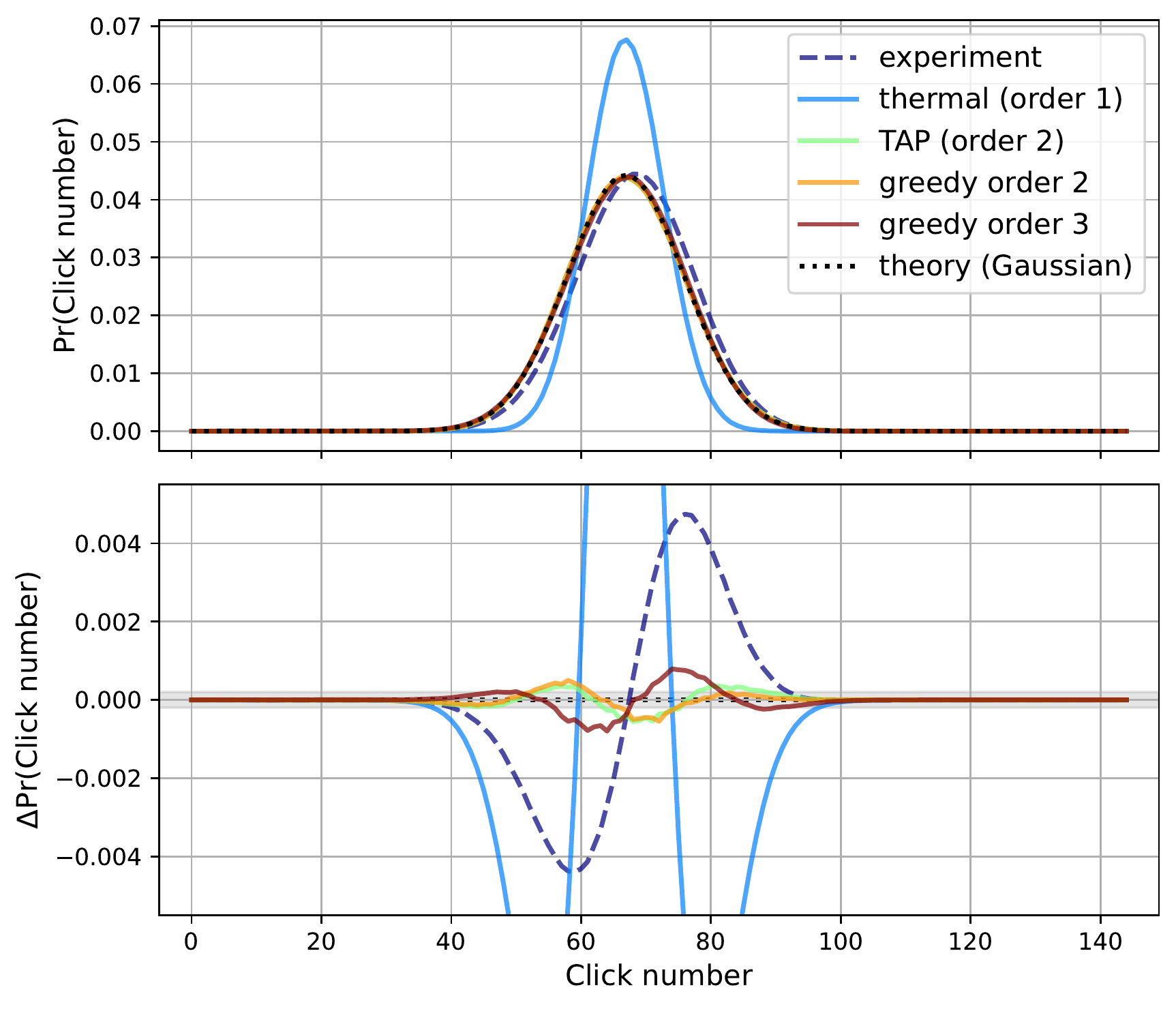}
\caption{\label{fig:hist_click_number} 
Distribution of the number of clicks  for the different samplers, i.e., experimental dataset \textbf{2.b.5} and mockup (upper panel), together with their difference with the theoretical prediction (lower panel).
The theoretical prediction for this distribution is plotted with a dotted line, and is approximated through a Gaussian; the error made by neglecting the third moment of the theoretical distribution is about $2\times 10^{-4}$ (shaded area in the bottom panel) and is smaller than the differences plotted.
See App.~\ref{app:moments} for details.
We observe that the thermal sampler performs worse than all others while, for this dataset, the higher order mockup samplers outperform the experiment.
} 
\end{figure}

%% file: results.tex
\begin{figure*}[t]
\centering
\includegraphics[width=2.00\columnwidth]{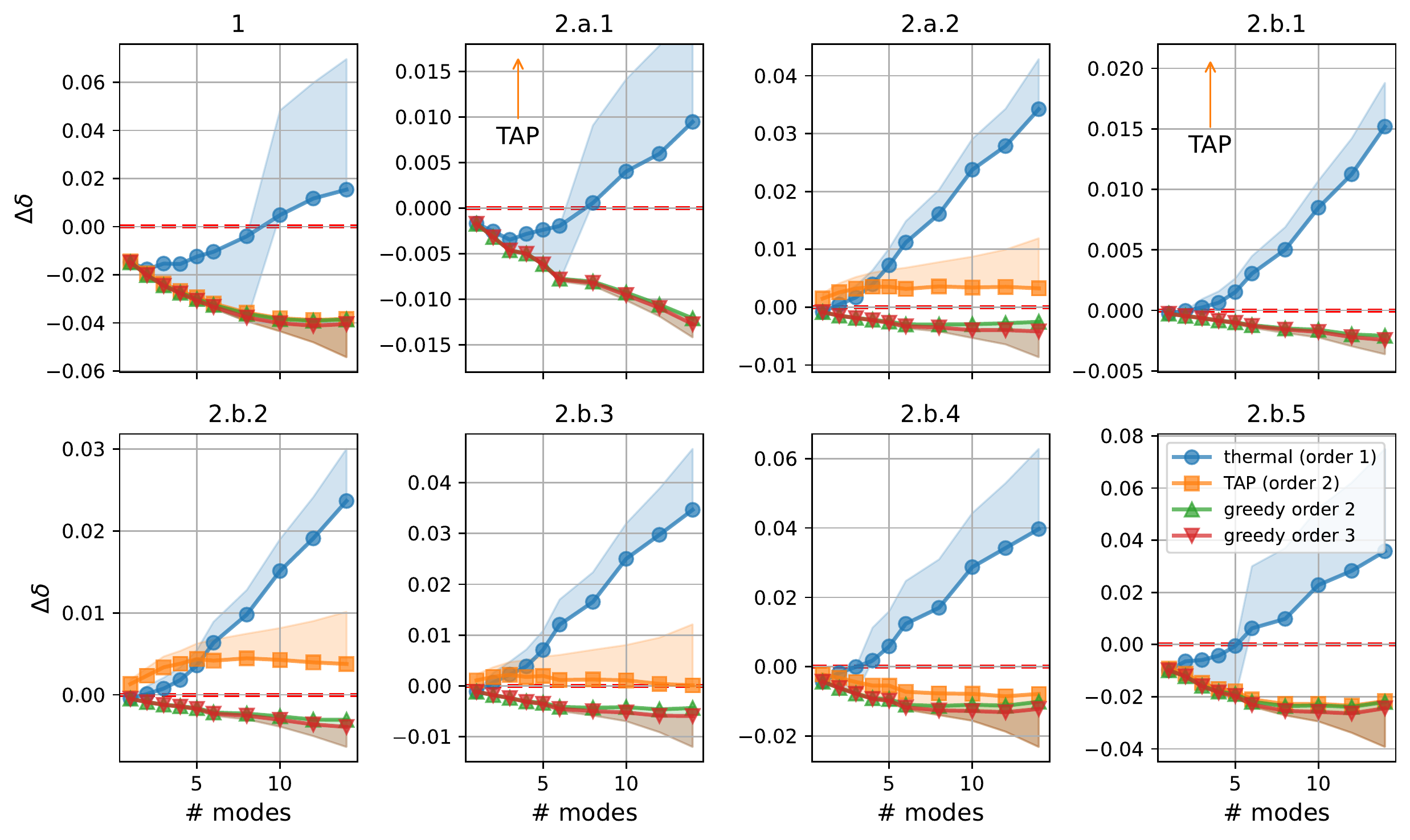}
\caption{\label{fig:variation_distance_difference} 
Total variation distance difference $\Delta \delta = \delta_m - \delta_e$ where $\delta_m$ is the distance between the ideal marginal distribution and a mockup, and $\delta_e$ is the distance between ideal and experiment. The title of each subplot denotes the dataset (see Table.~\ref{tab:datasets}). We consider marginal distributions from 1 to 14 modes.
Each point represents the average of $\Delta \delta$ over 100 randomly chosen subsets of modes. Error bars represent the empirical standard deviation.
For each subset of modes, the empirical probabilities  are computed with 10 million sampled bit strings.
A negative $\Delta \delta$ means the mockup sampler is sampling from a distribution with smaller total variation distance to the ideal distribution than the experiment, therefore outperforming the experiment according to this metric.
For a finite number of samples, the estimator of $\Delta \delta$ is biased towards 0. 
This bias becomes larger as the number of modes increases.
In order to have a converged estimate of $\Delta \delta$, 
an exponential number of samples would be needed.
Shaded areas represent lower and upper bounds of the actual value of $\Delta \delta$ (see App.~\ref{app:convergence} for details).
}
\end{figure*}

\begin{figure*}[t]
\centering
\includegraphics[width=2.00\columnwidth]{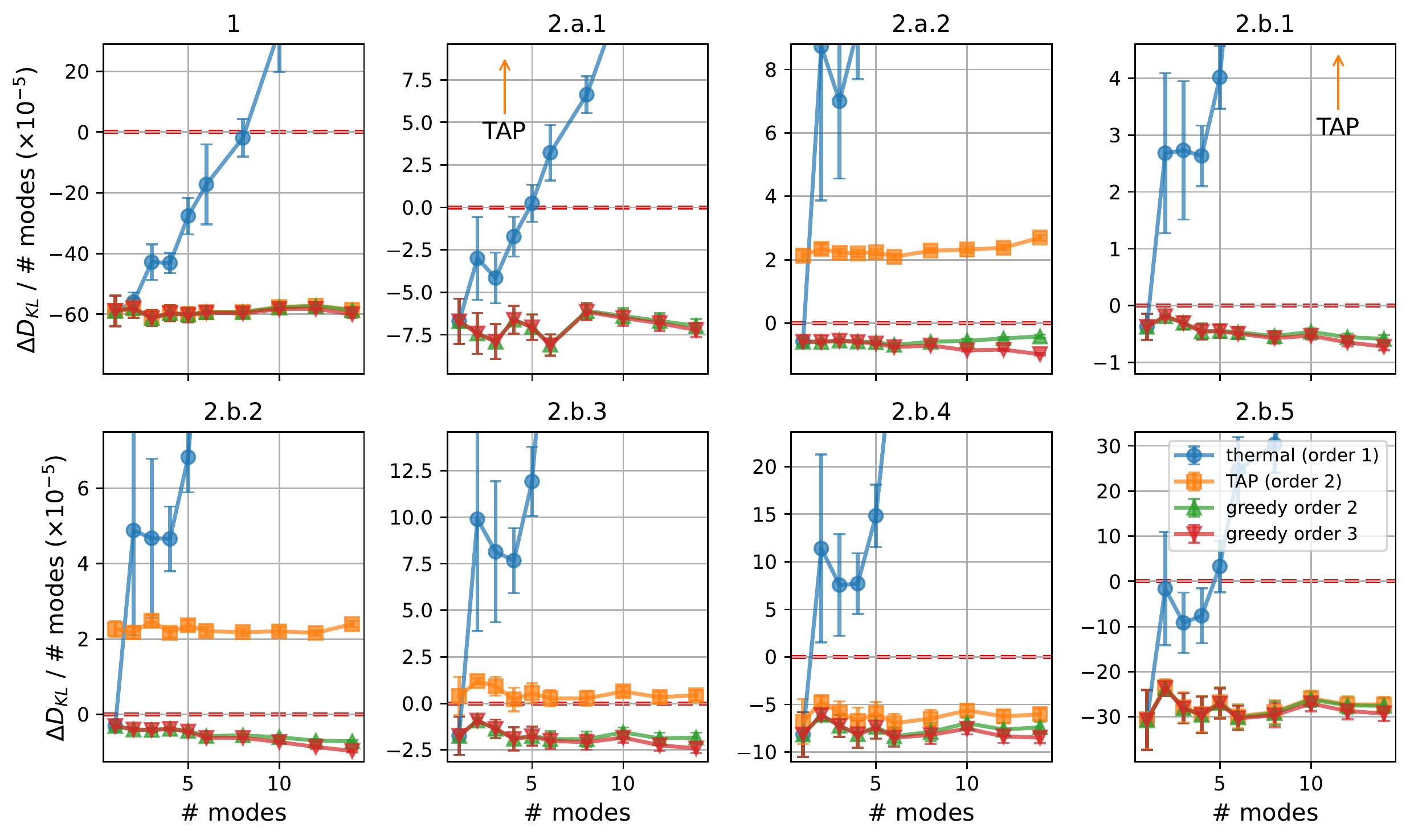}
\caption{\label{fig:kl_divergence_difference} 
Kullback-Leibler (KL) divergence difference per mode, $\Delta D_{\rm KL} / (\#\text{ modes})$, over marginal distributions of 1 through 14 modes between the mockup samplers and the experiment.
This figure is similar to Fig.~\ref{fig:variation_distance_difference}.
Each point represents the average of $\Delta D_{\rm KL} / (\#\text{ modes})$ over 100 randomly chosen subsets of modes of a certain size (or all 144 modes for the one mode case in experiment 2); error bars represent the standard error of the average.
For each subset of modes, the empirical probabilities, necessary to get $D_{\rm KL}$ (see main text), are computed over a set of 10 million sampled bit strings.
A negative $\Delta D_{\rm KL} / (\# \text{modes})$ means the mockup sampler is sampling from a distribution with smaller KL divergence, $D_{\rm KL}$, to the ideal distribution than the experiment, therefore outperforming the experiment.
} 
\end{figure*}

\begin{figure*}[t]
\centering
\includegraphics[width=2.00\columnwidth]{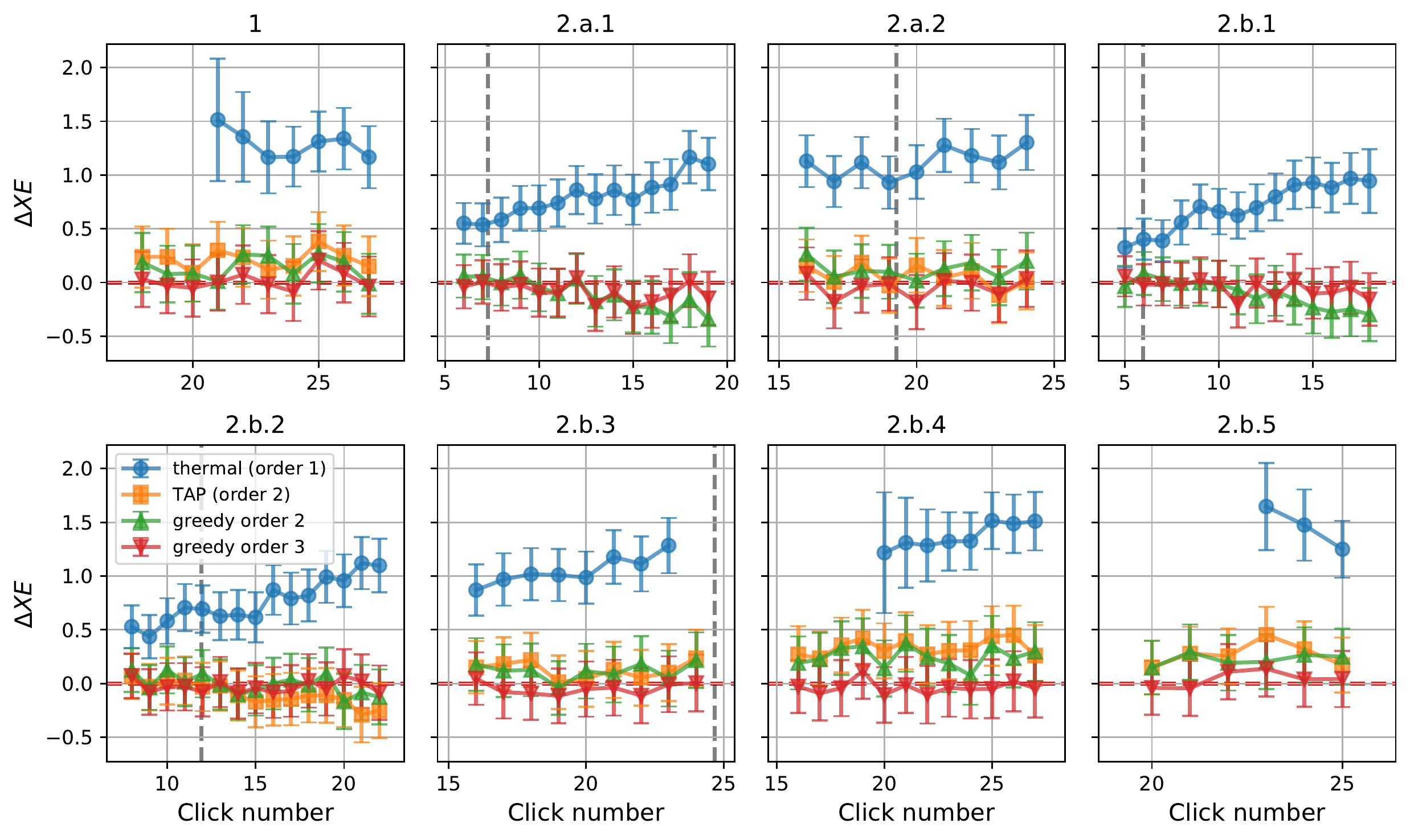}
\caption{\label{fig:xe_difference} 
Cross entropy difference, $\Delta {\rm XE}$, between mockup samples and experimental samples over bit strings with a fixed click number.
Markers represent the average over up to 1000 samples.
Error bars represent the propagated standard errors of the average XE of the experimental and the mockup samples.
For dataset \textbf{2.b.5}, we have marginalized the probability distribution over modes 1 though 90; this allows us to get enough bit strings with the studied click numbers.
Vertical dashed lines denote the average click number of the dataset.
Note that for datasets \textbf{2.a.1}, \textbf{2.b.1}, and \textbf{2.b.2} we are analyzing close to the full distribution, without the need to extrapolate to large click numbers. 
This is because the distributions of click number for these datasets have most of their mass over low click numbers (see Fig.~\ref{fig:click_number_dist}). 
While the thermal sampler is consistently worse than the experiment, the greedy sampler of order 3 performs similar to the experiment.
It is interesting to notice that the order 2 TAP sampler becomes better with larger power, i.e. over distributions with smaller single mode biases (1-mode marginals closer to $1 \over 2$).
With the TAP sampler, for datasets \textbf{2.a.1} and \textbf{2.b.1} we could not collect enough bit strings with the number of clicks studied.
Finally, all samplers seem to perform better with increasing click number.
} 
\end{figure*}

In this section we discuss the performance of the classical mockup samplers introduced in Section~\ref{sec:classical} as compared to the experimental quantum samplers of Refs.~\onlinecite{zhong_quantum_2020,zhong2021phase}.

We begin by comparing the Ursell functions (see App.~\ref{app:connected_correlations}) of the experiment and mockup samplers against their ideal values (see Fig.~\ref{fig:pearson}).
Mockup samplers of order $k$ exhibit only $k$th order Ursell functions by construction: the thermal sampler only shows correct 1st order Ursell functions, the 2nd order samplers show correct 1st and 2nd order correlations, with a fast decay beyond that order, and the third order greedy sampler shows correct Ursell functions up to 3rd order.
The 3rd order residual of the 2nd order samplers is almost identical, suggesting that the greedy sampler is performing similarly  to a maximum entropy sampler.
\footnote{Note that the $k$th order samplers have a $(k + 1)$th order residual.
It is not generally possible to construct a probability distribution with correlations of order $k$ or lower fixed to a given value, and higher order correlations equal to 0.}
Interestingly, the experiment shows correlation of the Ursell functions with their ideal values for all orders studied here (see Ref.~\onlinecite{zhong2021phase}), while the mockup samplers see a fast decay past the order of the sampler.

Next we consider the distribution of the number of clicks.
See Refs.~\onlinecite{drummond2021simulating,popova_cracking_2021} for recent studies on this quantity.
Fig.~\ref{fig:hist_click_number} shows the distribution Pr(click number) for the different mockup samplers for dataset \textbf{2.b.5}.
In addition, the dotted black curve represents a Gaussian (2nd moment) approximation to the theoretical distribution of the number of clicks, similar to the one introduced in Ref.~\onlinecite{popova_cracking_2021}.
The Gaussian approximation has a deviation of at most $\sim 2 \times 10^{-4}$ compared to the 3rd moment approximation.
We observe that the thermal sampler performs worse than all others while, for this dataset, the higher order mockup samplers outperform the experiment.

Our main result is a comparison of the total variation distance from the ideal distribution (ground truth) to the experiment and mockup samplers. On the one hand, it is not possible to estimate the total variation distance between for the samplers with a large mean number of clicks, both because the exact ideal probabilities are too hard to calculate, and also because it is not possible to produce enough samples of the experiment or mockup samplers to estimate the empirical probabilities. On the other hand, the ideal probabilities of the marginal distribution in a small subset of modes is easy to calculate (see App.~\ref{app:probabilities}) and the empirical marginal distribution for the same subset of modes can also be estimated. This is the approach that be follow to estimate the total variation distance. Fig.~\ref{fig:variation_distance_difference} plots the total variation distance difference $\Delta \delta = \delta_m - \delta_e$ where $\delta_m$ is the distance between the ideal marginal distribution and a mockup, and $\delta_e$ is the distance between ideal and experiment. The total variation distance between two distributions with probabilities $p(\bf z)$ and $q(\bf z)$ is
\begin{align}
    \delta = \frac 1 2 \sum_{\bf z}|p({\bf z}) - q({\bf z})| \;.
\end{align}
We consider marginal distributions from 1 to 14 modes.
\footnote{The estimation of $\Delta \delta$ using empirical probabilities is biased towards 0 and converges slowly in the number of samples used. This is more evident with a larger number of modes, due to the exponentially many probabilities and the fact that there is only a finite set of samples available to estimate them empirically.
The KL divergence shows much faster convergence with number of samples, which alleviates this problem.
See App.~\ref{app:convergence} for details.}
Note that mockup distributions are constructed using only ideal marginal probabilities with at most three modes.

We observe that while the thermal sampler is quickly outperformed by the experiment once a few modes are considered, the greedy samplers of order 2 and 3 outperform the experiment in total variation distance for marginal distributions.
Furthermore, these numerics suggest that the improvement $\Delta \delta$ either grows or stabilizes as a function of the number of modes.
Interestingly, on datasets \textbf{2.b}, the advantage in favor of the mockup samplers increases with increasing power, and therefore with increasing complexity of the ideal sampling problem. We also observe that experiment \textbf{1} has worse performance than experiment \textbf{2}.
The mean field Boltzmann sampler (TAP) performs similarly to the greedy sampler of order 2 for datasets \textbf{1} and \textbf{2.b.5}. However, the quality of this mean field sampler degrades for datasets with lower power, i.e., when the one-mode marginals are biased away from $1 \over 2$.

We obtain a similar result for the Kullback-Leibler (KL) divergence difference per mode, $\Delta D_{\rm KL} / (\#\text{ modes})$
\footnote{Note that the KL divergence is an extensive quantity.
Dividing by the number of modes makes it intensive and more convenient to analyze.}, over marginal distributions of 1 through 14 modes between the mockup samplers and the experiment.
The KL divergence between two distributions with probabilities $p(\mathbf{z})$ and $q(\mathbf{z})$ is
\begin{align}\label{eq:kl}
  D_{\rm KL}(p,q) = {\rm XE}(p,q) - H(p) = \sum_{\bf z} p({\bf z}) \log { p({\bf z}) \over q({\bf z}) } \;,
\end{align}
where XE is the cross-entropy and $H$ is the entropy. The cross-entropy is
\begin{align}\label{eq:xe}
  {\rm XE}(p, q) = -\sum_{\bf z} p({\bf z}) \log {q({\bf z}) } \;.
\end{align}

The KL divergence is a non-symmetric distance between two distributions. We choose $p(\mathbf{z})$ to be the mockup sampler probabilities and $q(\mathbf{z})$ to be the ideal probabilities. The cross-entropy XE$(p, q)$ can then be estimated for a larger number of clicks. This has been used in a similar context previously~\cite{boixo_characterizing_2018,arute_quantum_2019,wu_speedup_2019,zhu2021quantum,zhong_quantum_2020,zhong2021phase}. We analyze the XE at the end of the present section (see Fig.~\ref{fig:xe_difference}).

We observe in Fig.~\ref{fig:kl_divergence_difference} that the thermal sampler outperformed by the experiment once a few modes are considered, while the greedy samplers of order 2 and 3 outperform the experiment.
In general, the KL divergence difference per mode either gets wider or stabilizes as a function of the number of modes, and the 3rd order greedy sampler outperforms the 2nd order one.
In the case of datasets \textbf{2.a.2} and \textbf{2.b.1}, the 2nd order greedy sampler shows a difference with the experiment that shrinks with the number of modes, while the 3rd order sampler widens its difference with the number of modes; these two datasets are arguably the ones with highest quality experimental samples, as seen through other metrics too, such as the quality of the distribution of click numbers (see App.~\ref{app:moments} and Fig.~\ref{fig:moment_difference}).
As with the total variation distance of Fig.~\ref{fig:variation_distance_difference}, datasets \textbf{2.b} show the experiment degrades with power~\cite{zhong2021phase}. This is also true for experiment \textbf{1} as compared to experiment \textbf{2}.
Consistent with the total variation distance, we see that the mean field sampler TAP performs similarly to the greedy sampler of order 2 for datasets \textbf{1} and \textbf{2.b.5} and with lower quality  for datasets with lower power.

Let us now turn our attention to the cross entropy (XE) difference between experiment and mockup samples, see Eq.~\eqref{eq:xe}.
For a set of $n$ samples from $p({\bf z})$, $S_{\text{sampler}}$, we can estimate XE$(p, q)$ as
\begin{align}
    {\rm XE} &\simeq -\frac{1}{n} \log {\rm Pr}(S_{\rm sampler}) \\ &= -\frac{1}{n} \sum^n_{{i = 1}} \log\left[ q \left( \mathbf{z}_{{\rm sampler}, i} \right) \right] \text{,}
\end{align}
where $q(\bf z)$ is the ideal probability. 
This equation also corresponds to minus the average log-likelihood of the samples $S_{\rm sampler} = \{ \mathbf{z}_{{\rm sampler}, i}\}_{i = 1}^n$ with respect to the ideal probability distribution, $q$.
This estimator is convenient in cases where $n$ is small compared to the size of the sampling space, e.g., when samples consist of bit strings with a large number $N$ of modes. In this case it is intractable to compute empirical probabilities $p({\bf z})$. 

The tractability of computing the XE over larger systems (although still constrained by the exponentially hard computation of ideal probabilities) has made it a standard benchmark in both RCS~\cite{boixo_characterizing_2018,arute_quantum_2019,wu2021strong,zhu2021quantum} and GBS~\cite{zhong_quantum_2020,zhong2021phase}. In addition, in RCS, under fairly weak assumptions, the XE becomes an estimator of the system's fidelity~\cite{boixo_characterizing_2018,arute_quantum_2019}. More generally, note that the KL divergence is the difference between the cross-entropy XE and the sampler entropy, see Eq.~\ref{eq:kl}. Therefore, a sampler with low cross-entropy and high entropy will have small distance to the ideal distribution. Furthermore, if we assume that a noisy experimental sampler has entropy not lower than the ideal sampler, and the cross-entropy is close to the ideal entropy, then it has low total variation distance~\cite{bouland2019complexity}.
Nevertheless, a sampler with low cross-entropy and low entropy would be a bad sampler. For instance, a sampler that always outputs the bit string of all zeros, independently of the ideal GBS distribution, would have very low cross-entropy, because this bit string has relatively high probability. But it is obviously a bad sampler, and in particular it has very low entropy.~\footnote{Another example of a bad sampler with low entropy is the proposal of Ref.~\cite{pan_simulating_2021}, which in addition has exponential cost.
Indeed, the entropy of this sampler does not grow with system size.}

We show in Fig.~\ref{fig:xe_difference} the difference between the XE of the mockup samplers and that XE of the experiments, $\Delta {\rm XE}$, estimated over a set of up to 1000 samples, and for fixed click number sectors.
This method was used in Ref.~\onlinecite{zhong_quantum_2020} to compare several mockup samplers to the experiment.
Note that the click numbers studied on datasets \textbf{2.a.1}, \textbf{2.b.1}, and \textbf{2.b.2} cover virtually the full distribution (see Fig.~\ref{fig:click_number_dist}), which avoids the need to extrapolate to larger click numbers.
We observe that the thermal sampler has consistently larger XE than the experiment.
On the other hand, the 3rd order greedy sampler has similar XE as the experiment on all datasets.
The 2nd order samplers seem to have similar or slightly larger XE than the experiment (except for the large click number results mentioned above, where indeed they have smaller XE than the experiment).

Estimating the XE over sectors of fixed click number introduces two shortcomings.
First, the imperfect nature of the distribution of click number is not being considered.
As seen in Fig.~\ref{fig:click_number_dist}, the experiment can show non-negligible deviations from the ideal distribution of click number.
Second, the quality of the samples generated over each sector might not be consistent, as can be seen in datasets \textbf{2.a.1}, \textbf{2.b.1}, and \textbf{2.b.2}, where increasing the click number works in favor of the 2nd and 3rd order samplers, which show a decaying $\Delta$XE.

%% file: discussion.tex
In this work, we propose a family of classical methods for approximating experimental Gaussian boson sampling with a cost only quadratic in the number of modes.
We show that a 2nd order Boltzmann machine with parameters computed from a mean field approximation outperforms the experimental output from Refs.~\onlinecite{zhong_quantum_2020,zhong2021phase} over its hardest instances, as measured by total variation distance and KL divergence.
In addition, we introduce a heuristic, greedy method to generate samples with correct $k$th order marginal probabilities over the GBS output modes at a cost polynomial exponential in $k$ and polynomial in the number of modes. This method also outperforms the experimental output already at $k = 2$ and improves with higher $k$. The scaling of the distance to the ideal distribution with the order $k$ of the representation is an interesting open question.
The same methodology can be applied to other boson sampling proposals where marginal probabilities can be computed efficiently~\cite{ivanov2019complexity,clifford_classical_2017,clifford2018classical,aaronson2013bosonsampling,renema_marginal_2020}. We also review the relation between total variation distance, KL divergence and cross-entropy~\cite{bouland2019complexity}.

The $k$th order approximation reproduces Ursell functions only up to order $k$, with a cost exponential in $k$ and high precision, while the experiment exhibits higher order Ursell functions with lower precision. We do not attempt to produce here samples with similar high order Ursell functions to the experiment. Nevertheless, the theoretical computational hardness of boson sampling~\cite{aaronson2011computational}, GBS~\cite{hamilton_gaussian_2017,quesada_gaussian_2018,gupt_classical_2018,bjorklund_faster_2019,kruse_detailed_2019,zhong2019experimental,wu_speedup_2019,quesada_quadratic_2020,quesada_exact_2020,drummond2021simulating,li_benchmarking_2020}, IQP~\cite{bremner2011classical,bremner2016average} and RCS~\cite{boixo_characterizing_2018,aaronson2017complexity,bouland2019complexity,movassagh2019quantum,aaronson2019classical,bouland2021noise,kondo2021fine} is based on the difficulty of approximate sampling, for which total variation distance is a standard measure. The fact that a quadratic classical algorithm obtained a better approximation to the ideal distribution questions the computational hardness of the experiments in Refs.~\cite{zhong_quantum_2020,zhong2021phase}. This result does not apply to random circuit sampling where, in contrast to boson sampling, the only known polynomial approximation is the uniform  distribution over bit strings.
This highlights the advantages of a fully programmable quantum computer in increased computational capacity.

We show how to estimate the statistical distance between an experiment or a mockup distribution, and the ideal distribution. As experiments improve, in terms of input state preparation, photon-indistinguishability, photon-loss rate, system size, etc, the distance to the ideal distribution will improve. Higher order mockup distributions also have improved distance, with a cost exponential in the order. Understanding this distance quantitatively, in experiments and numerics, remains an open question.

%% file: appendix.tex
\section{GBS ground truth probabilities}
\label{app:probabilities}

The state at the output of a quantum linear optics experiment (in our case a GBS experiment) is described by the covariance matrix $\sigma$.
Detailed notes on how to compute $\sigma$ for the experiment of Ref.~\onlinecite{zhong_quantum_2020} are provided in App.~\ref{app:sigma}.
$\sigma$ is a matrix of size $2N \times 2N$, where $N$ is the number of output modes of the experiment.
Given an output bit string $\mathbf{z}$ with threshold detectors clicking on modes in the set $S$, its probability is computed as:
\begin{align}
\label{eq:probability_z}
    p(\mathbf{z}) = \frac{{\rm Tor}(O_S)}{\sqrt{\det(\sigma)}} \text{,}
\end{align}
where $O_S = \mathbb{1} - \left(\sigma^{-1} \right)_S$ and $A_S$ is the sub matrix of $A$ with rows $j$ and $j + N$ and columns $j$ and $j + N$, for all $j$ in the set $S$.
The so called Torontonian function ${\rm Tor}(A)$ is defined as:
\begin{align}
\label{torontonian}
    {\rm Tor}(A) = \sum_{Z \in P([|S|])} \frac{(-1)^{|Z|}}{\sqrt{\det \left( \mathbb{1} - A_Z \right)}} \text{,}
\end{align}
where $P([|S|])$ is the set of all $2^{|S|}$ subsets of $[|S|] = \{1, 2, \ldots, |S|\}$.
The cost of computing the determinant of a matrix of size $m \times m$ is $\mathcal{O}(m^3)$, and so the cost of computing the Torontonian is dominated by the exponentially many terms in the sum of Eq.~\eqref{torontonian}.
This cost scales as $\mathcal{O}(|S|^3 2^{|S|})$, i.e., exponentially in the number of clicks.

The partial trace of the quantum Gaussian state in a subset of modes $R$ has covariance matrix $\sigma_R$, which, similar to above, is the submatrix of $\sigma$ with rows and columns $j$ and $j + N$ for all j in the set $R$, and where $\sigma$ is of size $2N \times 2N$~\cite{serafini2017quantum}.
Computing marginalized probabilities is then also done with Eq.~\eqref{eq:probability_z}, starting with covariance matrix $\sigma_R$.
Note that the cost of computing a marginal probability on $k$ modes is exponential in the number of ones, which is at most $k$, and therefore efficient for $k$ fixed and small. 

GBS with threshold detectors, as well as the Torontonian function, was introduced in Ref.~\onlinecite{quesada_gaussian_2018}.
See that reference for a detailed derivation of the expressions presented above.
This appendix follows closely parts of that reference.

\section{Obtaining $\sigma$}
\label{app:sigma}

Refs.~\onlinecite{zhong_quantum_2020,zhong2021phase} do not provide the output matrix $\sigma$ of each dataset explicitly.
Instead, the data downloaded from \url{https://quantum.ustc.edu.cn/web/node/915} and \url{https://quantum.ustc.edu.cn/web/node/951} provides both the squeezing parameters $r_k$ of each dataset and the transformation matrix of the interferometer $T$.
There are only three transformation matrices: one for each set of datasets with fixed waist, i.e., \textbf{1}, \textbf{2.a} and \textbf{2.b} (see Table~\ref{tab:datasets} of the main text).
In this appendix we give a prescription for transforming this data into the covariance matrix $\sigma$ used in App.~\ref{app:probabilities}.
This appendix follows closely both Ref.~\onlinecite{quesada_gaussian_2018} and the Supplemental Material of Ref.~\onlinecite{zhong_quantum_2020}.

Matrix $\sigma$ is obtained from the expression:
\begin{align}
\label{eq:sigma}
\sigma = 
\mathbb{1} & - 
\frac{1}{2}
\begin{pmatrix}
T & 0 \\
0 & T^*
\end{pmatrix}
\begin{pmatrix}
T^\dagger & 0 \\
0 & T^T
\end{pmatrix} \nonumber \\
& +
\begin{pmatrix}
T & 0 \\
0 & T^*
\end{pmatrix}
\sigma_{\rm in}
\begin{pmatrix}
T^\dagger & 0 \\
0 & T^T
\end{pmatrix} \;,
\end{align}
where $T$ is a $N \times 50$ complex transformation matrix, with $N = 100$ in the experiment of Ref.~\onlinecite{zhong_quantum_2020} and $N = 144$ in the experiment of Ref.~\onlinecite{zhong2021phase}, $\sigma_{\rm in}$ is the covariance matrix describing the input state to the interferometer.
Note that matrix $T$ is not unitary, since it includes the effects of photon loss in the experiment, thus partially including noise in the ground truth of the experiment.
Note also that, as explained in the supplementary information of Ref.~\onlinecite{zhong_quantum_2020}, the phases of the squeezing parameters are absorbed in $T$, and $r_k$ are therefore real and positive.

We now turn our attention to obtaining $\sigma_{\rm in}$, which is simply the tensor product of 25 two-mode squeezed vacua:
\begin{align}
\label{eq:squeezed_vacuum}
\sigma_{\rm in} = S \sigma_{\rm vac} S^\dagger_{\rm TM} \;.
\end{align}
The vacuum covariance matrix is of the 25 pairs of input modes is $\sigma_{\rm vac} = {\mathbb{1} \over 2}$, of size $100 \times 100$.
The squeezing matrix $S$ is defined as
\begin{align}
\label{eq:squeezing_matrix}
S =
\left(
\begin{array}{ccc|ccc}
{\rm Ch}(r_1) & 0             & \dots  & {\rm Sh}(r_1) & 0             & \dots \\
0             & {\rm Ch}(r_2) & \dots  & 0             & {\rm Sh}(r_2) & \dots \\
\vdots        & \vdots        & \ddots & \vdots        & \vdots        & \ddots\\
\cline{1-6}
{\rm Sh}(r_1) & 0             & \dots  & {\rm Ch}(r_1) & 0             & \dots \\
0             & {\rm Sh}(r_2) & \dots  & 0             & {\rm Ch}(r_2) & \dots \\
\vdots        & \vdots        & \ddots & \vdots        & \vdots        & \ddots \\
\end{array}
\right)
\end{align}
with
\begin{align}
{\rm Ch}(r_k) &=
\begin{pmatrix}
\cosh(r_k) & 0 \\
0 & \cosh(r_k)
\end{pmatrix} \nonumber \\
{\rm Sh}(r_k) &=
\begin{pmatrix}
\sinh(r_k) & 0 \\
0 & \sinh(r_k)
\end{pmatrix} \;,
\end{align}
where $k = 1, 2, \ldots, 25$ and $S$ is of size $100 \times 100$.
Note that the squeezing phases do not appear in this expression, since they have been effectively absorbed in the transformation matrix $T$.
Importantly, note also that the order of rows and columns used in the definition of $S$ in Eq.~\eqref{eq:squeezing_matrix} is different from that one in the Supplemental Material of Ref.~\onlinecite{zhong_quantum_2020}.
Indeed, the order implied there was corrected in the text accompanying the downloadable data online, and this reordering of rows and columns is needed so the covariance matrix obtained, $\sigma$, is compatible with the transformation matrix $T$.
We believe that this confusion has led Ref.~\onlinecite{drummond2021simulating} to use the wrong experimental data in their analysis.

\section{Principle of maximum entropy}
\label{app:max_entropy}

We want to  find a probability distribution $p(\mathbf{z})$ such that its marginals up to order $k$ are equal to those of another probability distribution we want to approximate. The principle of maximum entropy tells us to choose the distribution with the largest entropy out of the family of distributions that satisfy these properties.

We write the constraints on the marginal probabilities in the form:
\begin{align}
\label{general_constraints}
    \sum_\mathbf{z} p(\mathbf{z}) f_\alpha(\mathbf{z}) = F_\alpha\;.
\end{align}
Let $p_{i_1,\ldots,i_l}$ denote the marginal probabilities that all the bits $\{i_1,\ldots,i_l\}$ are 1. We are interested in the constraints $F_\alpha \equiv p_{i^\alpha_1, \ldots i^\alpha_l}$, i.e., $F_1 \equiv p_1$, \ldots, $F_{m + 1} \equiv p_{12}$, etc. For a $k$th order approximation we have $l \le k$.
Because the variables are Boolean, this corresponds to $f_\alpha(\mathbf{z}) \equiv z_{i^\alpha_1} \ldots z_{i^\alpha_l}$. 
More explicitly: $f_1(\mathbf{z}) \equiv z_1$, $f_2(\mathbf{z}) \equiv z_2$, \ldots, $f(\mathbf{z})_{m + 1} \equiv z_1 z_2$, $f_{m + 2}(\mathbf{z}) \equiv z_1 z_3$, etc.

The general solution for the maximum entropy distribution obeying constraints as in Eq.~\ref{general_constraints} is
\begin{align}
\label{eq:general_solution}
    p(\mathbf{z}) = \frac{1}{Z} \exp[\sum_{\alpha = 1}^n \lambda_\alpha f_\alpha(\mathbf{z})]\;.
\end{align}
In our case this gives a Boltzmann machine:
\begin{align}
\label{eq:bma}
    p(\mathbf{z}) = \frac{1}{Z} \exp \left[\sum_{a} \lambda_a z_a + \sum_{a < b} \lambda_{a, b} z_a z_b \right. \nonumber \\
    \left.+ \sum_{a < b < c} \lambda_{a, b, c} z_a z_b z_c + \ldots \right]\;.
\end{align}

Note that the constraints used to derive Eq.~\ref{eq:bma} completely determine any $l$-bit marginal. Indeed the $2^l - 1$ degrees of freedom of an $l$-bit marginal probability distribution are given by the set of 1-bit marginal probabilities, $\{p_{i_1}, \ldots, p_{i_l}\}$, together with the 2-bit marginal probabilities $\{p_{i_m, i_n}\}_{1 \le m < n \le l}$, together with the 3-bit marginal probabilities, etc.

\begin{figure}[t]
\centering
\includegraphics[width=1.00\columnwidth]{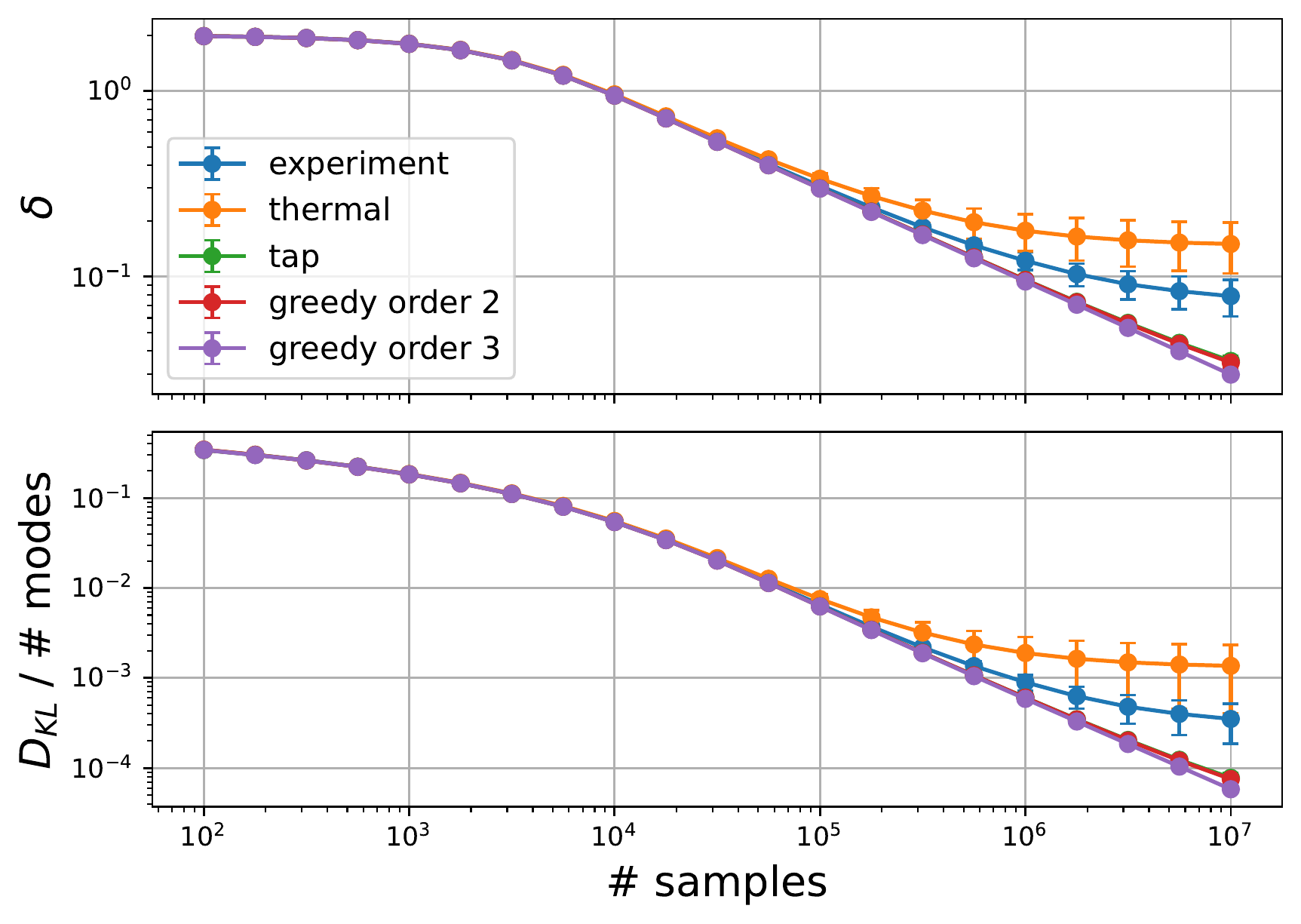}
\caption{\label{fig:tvd_and_kl_vs_samples} 
Estimates of the total variation distance $\delta$ (top) and the KL divergence $D_{\rm KL}$ as a function of the number of samples for the different samplers, i.e., experimental and mockup, over subsystems of 14 modes on dataset \textbf{2.b.5}.
Markers represent averages over 100 randomly chosen subsystems of 14 modes and errorbars represent standard deviations.
We can see that estimates are biased towards larger distances.
All estimates are not converged, although larger distances are closer to convergence than small ones.
} 
\end{figure}

\begin{figure*}[t]
\centering
\includegraphics[width=2.00\columnwidth]{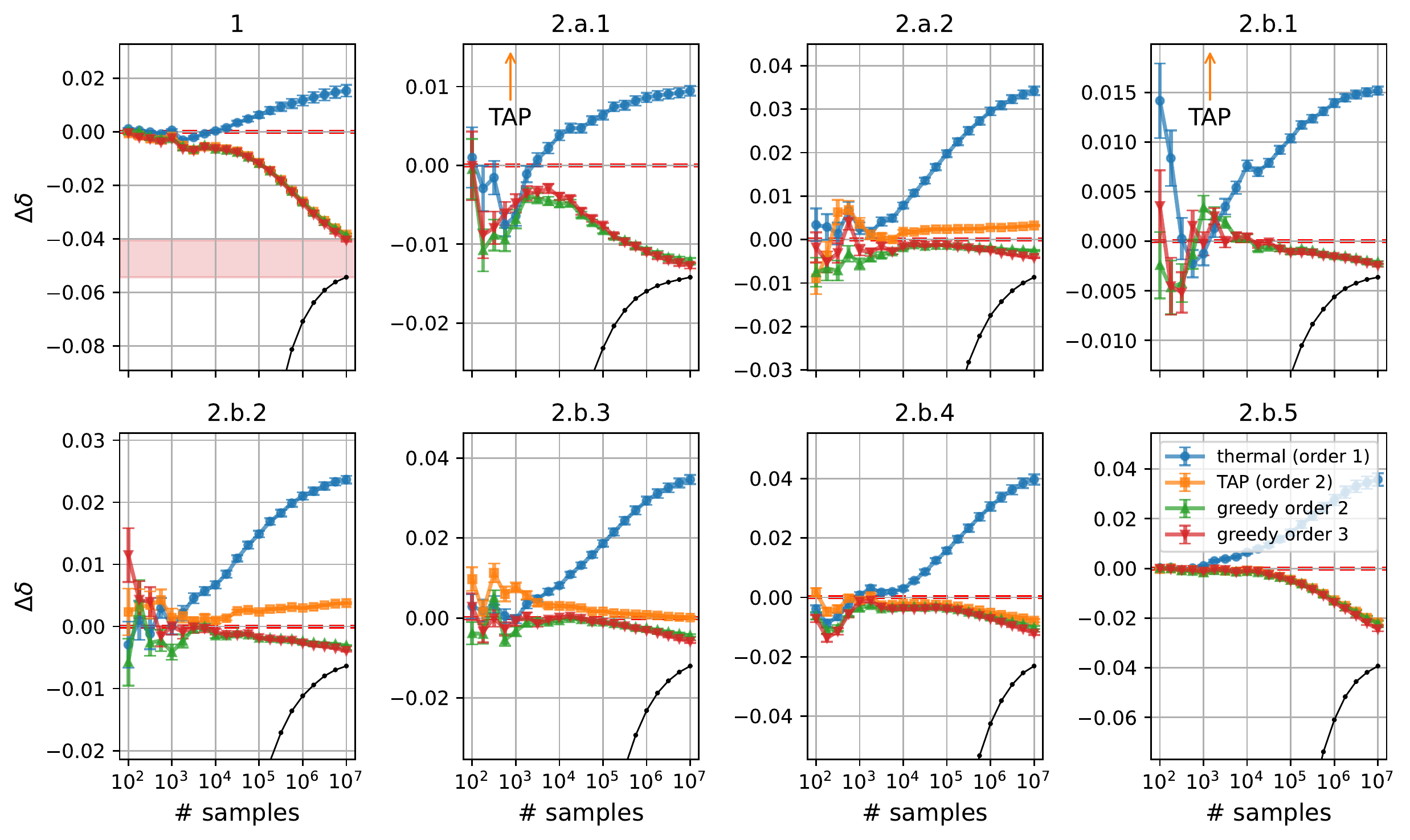}
\caption{\label{fig:tvd_vs_samples} 
Estimates of the total variation distance difference $\Delta \delta = \delta_m - \delta_e$ as a function of the number of samples for mockup samplers on subsystems of 14 modes for all datasets.
Markers represent averages over 100 randomly chosen subsystems and error bars represent standard errors.
Solid black lines represent $-\delta_e$, which is used as a lower bound of $\Delta \delta$ for mockup samplers with negative $\Delta \delta$.
Prior to convergence estimates are biased towards 0 and can be used as upper (lower) bounds of $\Delta \delta$.
As an example, the top left panel shows with a shaded area the lower and upper bounds estimated with 10 million samples for the 3rd order greedy sampler.
} 
\end{figure*}

\begin{figure*}[t]
\centering
\includegraphics[width=2.00\columnwidth]{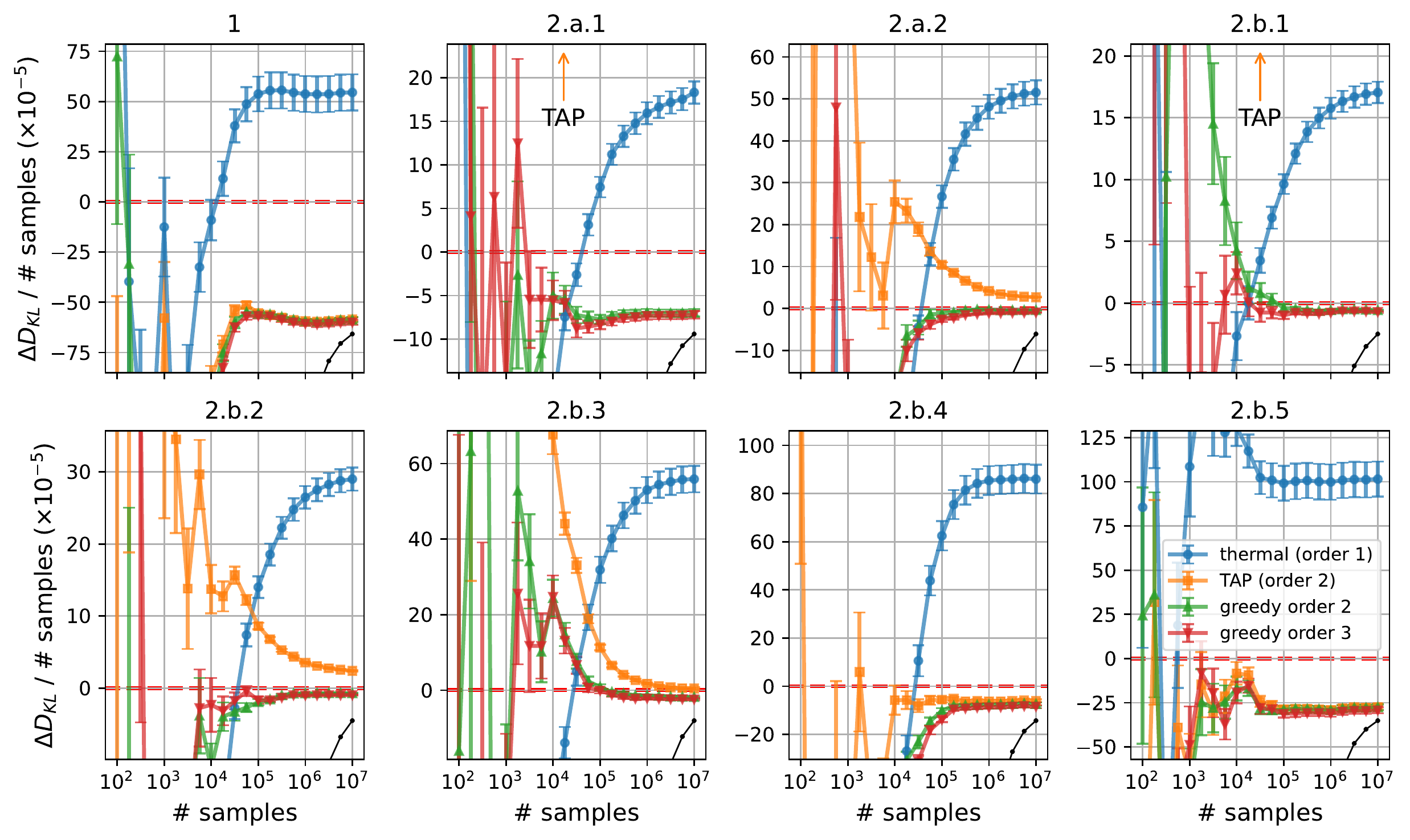}
\caption{\label{fig:kl_vs_samples} 
Estimates of the KL divergence difference per mode $\Delta D_{\rm KL} / ({\rm \#\; modes})$ as a function of the number of samples for mockup samplers on subsystems of 14 modes for all datasets. 
Markers represent averages over 100 randomly chosen subsystems and error bars represent standard deviations.
Solid black lines represent $\delta_e$, which are shown for completeness, but are not used to estimate a lower bound for $\Delta D_{\rm KL}$.
We can see that $\Delta D_{\rm KL}$ converges with a moderate number of samples and has therefore been estimated with high precision.
} 
\end{figure*}

\begin{figure}[t]
\centering
\includegraphics[width=1.00\columnwidth]{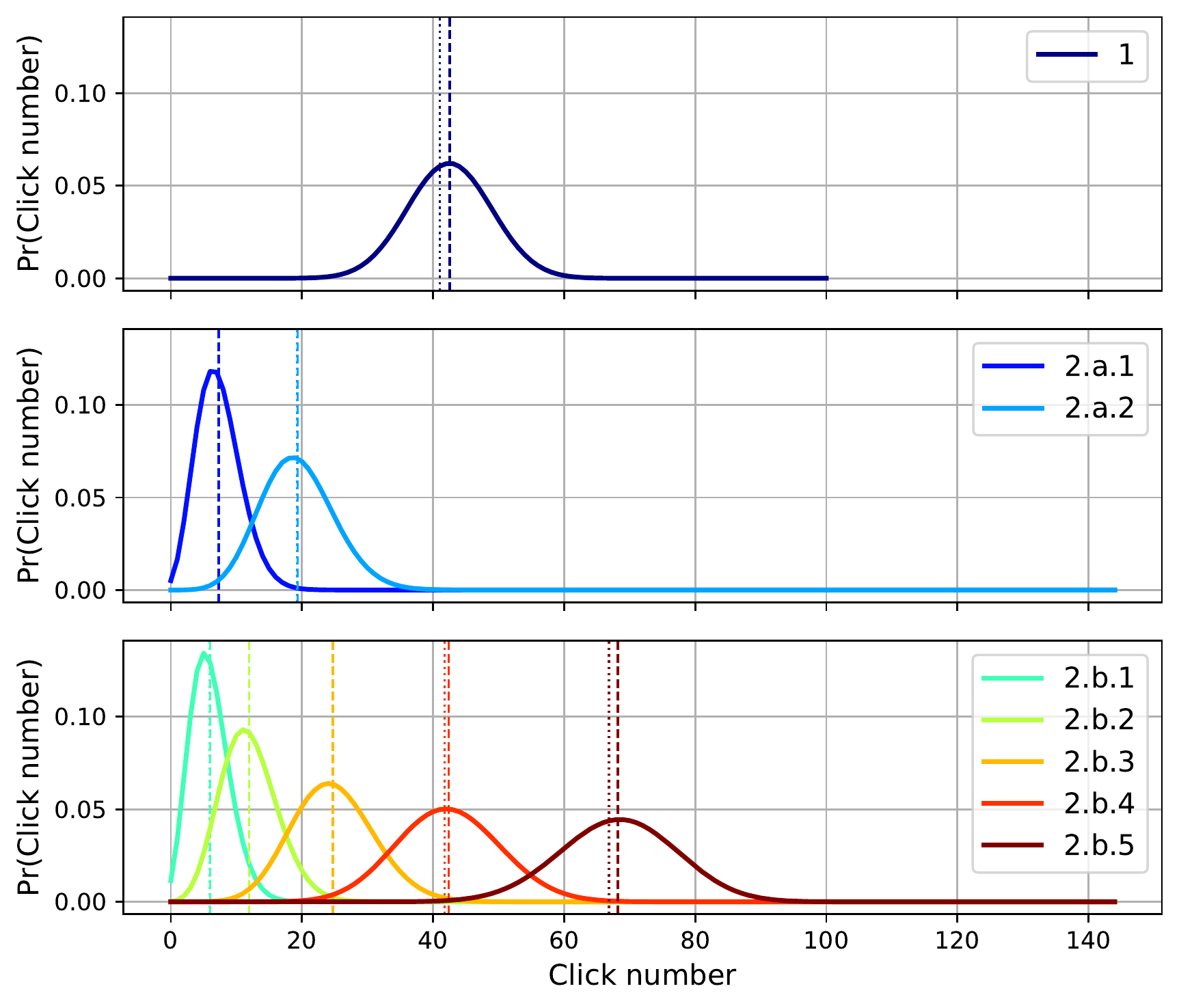}
\caption{\label{fig:click_number_dist} 
Experimental distributions of the click number of the sampled bit strings.
See Table~\ref{tab:datasets} for a characterization of the different datasets.
Dashed vertical lines denote the empirical average of the distributions, while dotted vertical lines denote the average of the ideal distributions of click numbers. The
ideal distributions are not shown.
Note that the disagreement in these averages grows with the power of the experiment, i.e., as the experiment detects more clicks.
This is an indication of the degradation of the quality of the experiment with its complexity.
} 
\end{figure}

\begin{figure}[t]
\centering
\includegraphics[width=1.00\columnwidth]{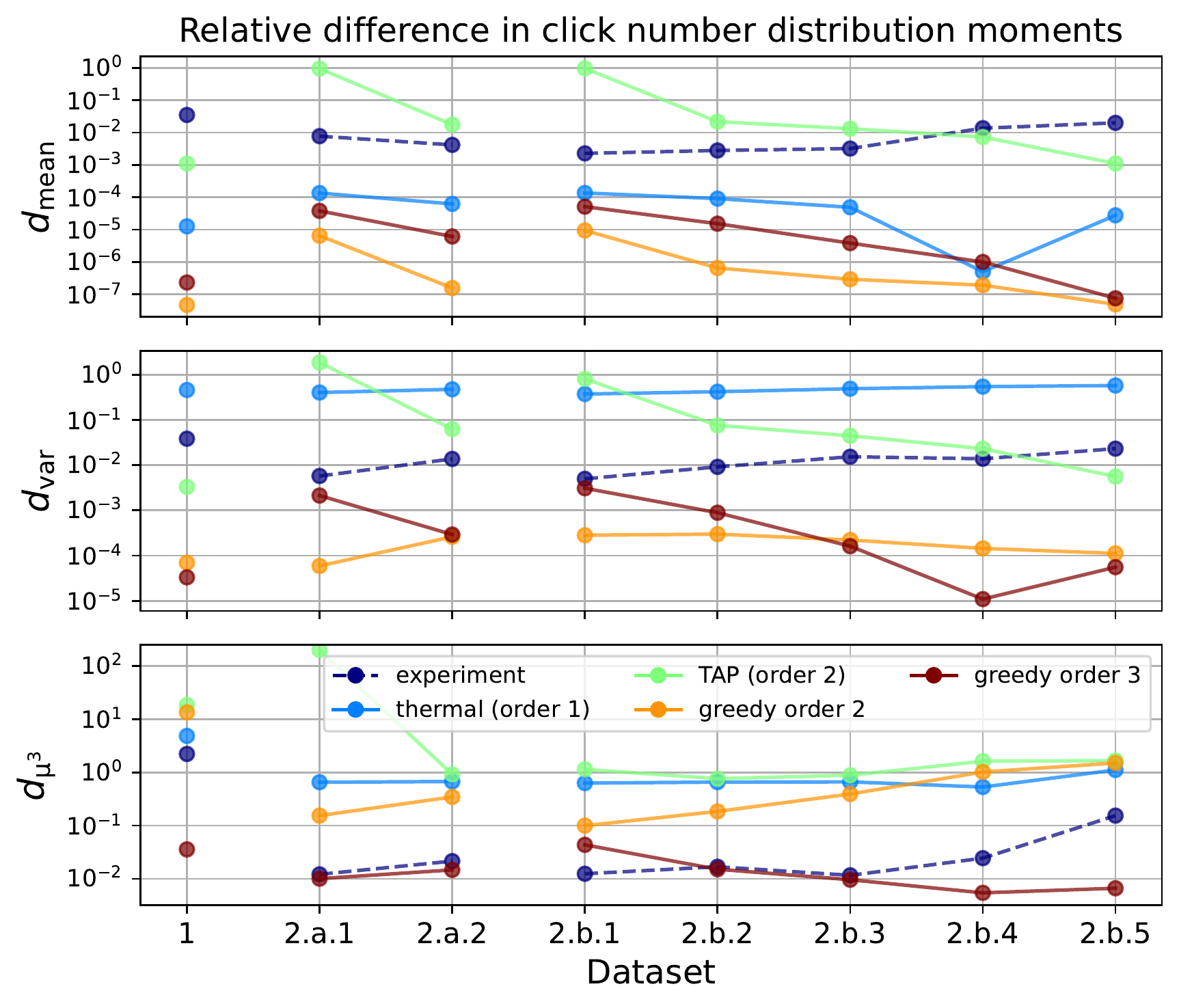}
\caption{\label{fig:moment_difference} 
Absolute value of the relative difference between the ideal moments of the distribution of click number and their empirical counterparts for the different samplers.
We study the mean, the variance, and the third moment with respect to the mean, $\mu^3$.
See App.~\ref{app:moments} for details on how to compute the ideal moments.
} 
\end{figure}

\section{Ursell functions}
\label{app:connected_correlations}

Given a single mode $a$ of a GBS experiment, we define its 1st order ``correlation'' $d^a$ as $d^a = \mathbbm{E}[z_a] - \frac{1}{2}$, i.e., the difference between its marginal probability of click and the uniform distribution marginal probability.
Note that this definition is different from the standard one, which would not include the subtraction of $\frac{1}{2}$.

For order $k > 1$, the $k$-mode Ursell function over $k$ modes $\vec{a} = a_1, \ldots, a_k$ is the difference between $\mathbbm{E}[z_{a_1} z_{a_2} \ldots z_{a_k}])$ and the sum, over all partitions of the modes in $\vec{a}$, of the product of Ursell functions of the subsets of each partition (or the one-mode probability for subsets with only one mode):
\begin{align}
\label{eq:connected_correlations}
d^{\vec{a}} = p^{\vec{a}}(z_{a_1}, \ldots, z_{a_k}) - \sum_{\mathcal{A} \in P[\vec{a}]} \prod_{\vec{\alpha} \in \mathcal{A}} \bar{d}^{\vec{\alpha}} \;,
\end{align}
where $P[\vec{a}]$ is the set of all partitions of the set of modes $\vec{a}$, $\vec{\alpha}$ iterates over all subsets in partition $\mathcal{A}$, and $\bar{d}^{\vec{\alpha}} = d^{\vec{\alpha}}$ if set $\vec{\alpha}$ has more than one mode, and $\bar{d}^{\vec{\alpha}} = \mathbbm{E}[z_{\alpha_1}]$ if set $\vec{\alpha}$ has only one mode, namely $\alpha_1$.
For illustration purposes, let us explicitly write down the expression for the 3rd order Ursell functions $d^{\vec{a}} = d^{a_1, a_2, a_3}$:
\begin{align}
\label{eq:connected_correlation_example}
d^{\vec{a}} = &\mathbbm{E}[z_{a_1} z_{a_2} z_{a_3}] \nonumber \\
&- \mathbbm{E}[z_{a_1}] \mathbbm{E}[z_{a_2}] \mathbbm{E}[z_{a_3}]  \nonumber \\
&- \mathbbm{E}[z_{a_1}] d^{a_2, a_3} - \mathbbm{E}[z_{a_2}] d^{a_1, a_3}  - \mathbbm{E}[z_{a_3}] d^{a_1, a_2} \;.
\end{align}

Ursell functions, which are also known as connected correlations, were originally introduced by Ursell through the equivalent definition~\cite{ursell1927evaluation} (see also Refs.~\onlinecite{duneau1973decrease,Walschaers2018,phillips2019benchmarking,zhong2021phase}):
\begin{align}
\label{eq:connected_correlations_ursell}
d^{a_1, \ldots, a_k} = &\left. \frac{\partial}{\partial r_1} \ldots \frac{\partial}{\partial r_k} \right. \nonumber \\
&\left. \log \left\{ \mathbbm{E} \left[ \exp(\sum_{i = 1, \ldots, k}{r_i z_{a_i}}) \right] \right\} \right\vert_{\vec{r} = 0} \;.
\end{align}

\section{Gibbs sampling from a Boltzmann machine}
\label{app:gibbs}

Given a BM as in Eq.~\eqref{eq:bm} or Eq.~\eqref{eq:ising} of the main text, it is simple to perform Gibbs sampling from it.
In practice we have used the spin representation of the BM (i.e., Eq.~\eqref{eq:ising}) in our implementation with parameters gotten from the mean field TAP approximation, so let us focus on that representation in this appendix.
Gibbs sampling is a Markov Chain Monte Carlo (MCMC) algorithm to sample from a distribution whose conditional probabilities of one variable conditioned on all others are known.
First, we choose a random bit string.
We then choose a mode, say mode $a$, and choose $s_a$ to take value $-1$ or 1 with probability equal to $p(s_a | \{s_i \}_{i \neq a})$.
We iterate over all modes sequentially and, at each iteration, choose the corresponding bit with its probability conditioned on all other modes being set to their current value.
This MCMC algorithm requires a burn-in period, in order to converge, and a thinning period, in order to reduce correlations between consecutive samples.
In practice, we use a burn-in period of 15000 and a thinning period of 900. Note that we have the number of modes is 100 in experiment \textbf{1} and 144 in experiment \textbf{2}.

The probability of mode $a$ taking value $s_a = -1$ conditioned on all others can be computed efficiently from Eq.~\eqref{eq:ising}:
\begin{align}
\label{eq:conditional_p}
    p(s_a = -1 | \{s_i \}_{i \neq a}) = \frac{1}{1 + \exp{ 2 h_a + 2 \sum_{i \neq a} J_{ai} s_i }} \; ,
\end{align}
The conditional probability of $s_a = +1$ is simply $p(+1 | \{s_i \}_{i \neq a}) = 1 - p(-1 | \{s_i \}_{i \neq a})$.
Note that evaluating the partition function $Z$, which normalizes the expression in Eq.~\eqref{eq:ising}, is not needed in Eq.~\eqref{eq:conditional_p}.
Eq.~\ref{eq:conditional_p} is evaluated in time linear in the number of modes $N$.
Since we need to iterate over all $N$ modes repeatedly, the overall cost per sample scales as $N^2$, and the time complexity to generate $L$ samples is $\mathcal{O}(N^2 L)$

\section{HOG rate and $\Delta$XE}
\label{app:hog}

In order to compare experimental samples $S_{\rm experiment}$ to mockup samples $S_{\rm mockup}$, the authors of Ref.~\onlinecite{zhong_quantum_2020} define the HOG rate as the ratio
\begin{align}
\label{hog}
    r_{\rm HOG} = \frac{{\rm Pr}\left(S_{\rm experiment} \right)}{{\rm Pr}\left(S_{\rm experiment} \right) + {\rm Pr}\left(S_{\rm mockup} \right)} \text{,}
\end{align}
where the probability ${\rm Pr}\left(S\right)$ is defined by the ground truth as in App.~\ref{app:probabilities}.
We can rewrite this expression as:
\begin{align}
\label{hog_XE_difference}
    r_{\rm HOG} = \frac{1}{1 + e^{ n \left( {\rm XE}_{\rm experiment} - {\rm XE}_{\rm mockup} \right) }} = \frac{1}{1 + e^{ -n\Delta {\rm XE} }}\;,
\end{align}
where the cross-entropy XE is defined in Section~\ref{sec:results} of the main text.
For large $n$ ($n \approx 1000$ in practice) $r_{\rm HOG}$ gives either 0 or 1, depending on whether the sign of the XE difference is negative or positive, respectively.
Given the fluctuations of the sign of $\Delta {\rm XE}$ for all mockup samplers other than the thermal (Fig.\ref{fig:xe_difference} of the main text) we choose to show the raw data instead of the HOG rate.
Note that, given that the cost of computing single bit string probabilities is exponential in their click number, $\Delta {\rm XE}$ is estimated over subspaces of a fixed click number, as is the HOG rate in Ref.~\cite{zhong_quantum_2020}.

\section{Estimates and bounds on the total variation distance and KL divergence with a finite number of samples}
\label{app:convergence}

Both the total variation distance $\delta$ and the KL divergence $D_{\rm KL}$ suffer from a bias when estimated from empirical probability distributions.
In this section we analyze this effect on the estimation of $\Delta \delta$ and $\Delta D_{\rm KL}$ in the main text.
We will see that, on the one hand, for the largest subsystems studied (14 modes) $\Delta \delta$ is far from converged using 10 million samples.
We can however estimate lower and upper bounds for this quantity.
On the other hand, $\Delta D_{\rm KL}$ converges quickly as a function of the number of samples, and the results presented in Fig.~\ref{fig:kl_divergence_difference} are precise.

Fig.~\ref{fig:tvd_and_kl_vs_samples} shows the estimate of $\delta$ and $\Delta D_{\rm KL}$ as a function of the number of samples for the different samplers and for subsystems of 14 modes on dataset \textbf{2.b.5}.
Both quantities are overestimated when using a small number of samples.
Although no sampler has converged with 10 million samples, samplers with a larger distance converge faster than those with a smaller distance.

We now look at the convergence of difference between distances, i.e., $\Delta \delta$ and $\Delta D_{\rm KL}$, as a function of the number of samples averaged over subsystems of 14 modes.
Fig.~\ref{fig:tvd_vs_samples} shows the estimates of the total variation distance difference $\Delta \delta = \delta_m - \delta_e$, where $\delta_m$ is the total variation distance between the ideal marginal distribution and that one a of a set of mockup samples, and $\delta_e$ is the distance between the ideal and the experiment.
We can see that $\Delta \delta$ is far from converged with 10 million samples.
We find that estimates of $\Delta \delta$ are biased towards 0, which allows these estimates to be used as lower (upper) bounds of $\Delta \delta$ when this quantity is negative (positive).
We therefore estimate an upper (lower) bound of $\Delta \delta$ as $-\delta_e$ ($\delta_m$).
The black line on each panel shows the estimates of $-\delta_e$, which serve as a lower bound to $\Delta \delta$ for curves with $\Delta \delta < 0$.
As an example, the top left panel shows the area in between the lower and upper bounds of $\Delta \delta$ of the 3rd order greedy sampler (shaded). 
These bounds are used in Fig.~\ref{fig:variation_distance_difference} in the main text.

Fig.~\ref{fig:kl_vs_samples} shows the estimates of $\Delta D_{\rm KL}$ per mode as a function of the number of samples averaged over subsystems of 14 modes.
As opposed to $\Delta \delta$, $\Delta D_{\rm KL}$ converges on most cases to a precise value with a modest number of samples.
This is certainly the case for the curves with $\Delta D_{\rm KL} < 0$.
For completeness, the black line shows $-D_{\rm KL}$ of the experimental data.
Given the convergence of $\Delta D_{\rm KL}$, we do not use the black line as a lower bound of this quantity in Fig.~\ref{fig:kl_divergence_difference} of the main text.

\section{Click number distributions and their moments}
\label{app:moments}

In this appendix we study the distributions of click number of the experiment and mockup samplers.
The empirical distributions of click number of the experiments are shown in Fig.~\ref{fig:click_number_dist}.
We can see that both reducing the waist and increasing the power of the pump increases the overall click number in the output.

In order to study the quality of the experimental and mockup data when compared to the ideal distributions of click number, we now proceed to analyze the empirical and theoretical values of the low order moments of these distributions, for the experiment and mockup samplers.
We first derive expressions for the calculation of the ideal moments of the click number distributions.

The $k$th moment of the distribution of click number with respect to its mean can be computed from the set of all $k$th and lower order marginals of the theoretical distribution $p(\mathbf{z})$.
This is done by first writing the click number operator as $\hat{Z} = \sum_a z_a$ and then writing its $k$th order moment as
\begin{align}
\label{eq:kth_moment}
    \mu^k &= \mathbbm{E}\left[ \left(\sum_a z_a - \mu^1 \right)^k\right] \;,
\end{align}
where $\mu^1$ is the mean click number.
The binomial in Eq.~\eqref{eq:kth_moment} can be expanded in terms of the moments of order $k^\prime \leq k$ \emph{with respect to 0}, which are computed through 
\begin{align}
\label{eq:kth_moment_to_0}
    \mathbbm{E}&\left[ \left(\sum_a z_a \right)^{k^\prime}\right] = \sum_{\mathbf{z}} \left[ \sum_{a_1, a_2, \ldots, a_{k^\prime}} \left(z_{a_1} z_{a_2} \ldots z_{a_{k^\prime}} \right) p(\mathbf{z}) \right] \nonumber \\
    &= \sum_{a_1, a_2, \ldots, a_{k^\prime}} p^{a_1, a_2, \ldots, a_{k^\prime}}(11 \ldots 1) \nonumber \\
    &= \sum_{l = 1}^{k^\prime} t({k^\prime}, l) \sum_{a_1 < a_2 < \ldots < a_l} p^{a_1, a_2, \ldots, a_l}(11 \ldots 1) \text{,}
\end{align}
where $p^{a_1, a_2, \ldots, a_l}(z_{a_1} z_{a_2} \ldots z_{a_l})$ is the marginal probability of the $k$-bit string $z_{a_1} z_{a_2} \ldots z_{a_l}$ over modes $(a_1, a_2, \ldots a_l)$ and the combinatorial factor $t(k^\prime, l)$ is equal to
\begin{align}
\label{combinatorial}
    t(k^\prime, l) = \sum_{n_1 + n_2 + \ldots + n_l = k^\prime} \binom{k^\prime}{n_1, n_2, \ldots, n_l} \text{,}
\end{align}
where $n_1, \ldots, n_l > 0$.
We have used the fact that the sum over all bit strings of the product $(z_{a_1} z_{a_2} \ldots z_{a_l}) p(\mathbf{z})$ is effectively summing over all configurations of the bits that are not in the set $\{ a_1, \ldots, a_l \}$ conditioned to the bits in the set being all equal to 1.
This is equal to the marginal probability of the bit string of all 1s over the set of modes $\{ a_1, \ldots, a_l\}$.
This computation is similar to the one introduced in Ref.~\onlinecite{popova_cracking_2021}.
Note that Ref.~\onlinecite{drummond2021simulating} derived expressions to calculate the theoretical click number distribution exactly, although we do not make use of them here.

Fig.~\ref{fig:moment_difference} shows the relative difference between the empirical moments of the distribution of click number of the experimental and mockup samplers and their theoretical, ideal values, for all datasets and up to order 3.
In general, we see that a $k$th order sampler only approximates moments of the click number distribution up to order $k$, as expected from Eqs.~\eqref{eq:kth_moment} and~\eqref{eq:kth_moment_to_0}.
In addition, the relative difference of the moments (of order up to the order of the sampler) usually becomes smaller as the click number distributions shift towards larger click numbers, i.e., as the power of the experiment increases.
This is in contrast to the experiment, for which the relative difference with the theoretical values of the moments becomes larger as the power increases, consistent with the degradation in the quality of the experimental output~\cite{zhong2021phase}.\footnote{Interestingly, the relative difference of the mean improves from dataset \text{2.a.1} to \text{2.a.2}, consistent with Figs.~\ref{fig:variation_distance_difference} and~\ref{fig:kl_divergence_difference} of the main text.}
For mockup samplers other than TAP we see better performance (for orders smaller or equal to the order of the sampler) than the experiment.
The TAP mean field sampler outperforms the experiment in this metric over the datasets with the largest click numbers, consistent with $\Delta \delta$ and $D_{\rm KL} / \# {\rm \ modes}$ (see Figs.~\ref{fig:variation_distance_difference} and~\ref{fig:kl_divergence_difference} of the main text).
While $k$th order samplers only approximate moments of order $\leq k$, we expect the experiment to approximate higher order moments of this distribution, albeit with degrading quality as the order increases.
Finally, it is interesting to notice that the data of experiment \textbf{2} shows better performance than experiment \textbf{1}.

Fig.~\ref{fig:hist_click_number} of the main text includes an approximation to the theoretical distribution of click number.
We obtain this curve by finding the constants in an exponential function of the form $\exp(A + Bx + C x^2)$, where $x$ is the click number, such that its first two moments match the ideal ones.
For the data of Fig.~\ref{fig:hist_click_number}, we find that a 3rd order approximation to this distribution, i.e., using an ansatz of the form $\exp(A + Bx + Cx^2 + D x^3)$ and the first three moments, results only in a negligible correction.
This correction is smaller than the differences found with the empirical distributions of the samplers.
These approximations to the click number distribution are similar to those introduced in Ref.~\onlinecite{popova_cracking_2021}.
Incidentally, these exponential functions correspond to the maximum entropy solution for the click number distribution with constrained low order moments.
Note that Ref.~\onlinecite{drummond2021simulating} recently introduced a procedure to calculate the click number distribution exactly.